\documentclass[9pt,twocolumn,twoside]{pnas-new}
\templatetype{pnasresearcharticle}

\usepackage{xcolor}

\setboolean{displaywatermark}{false}
\makeatletter
\fancypagestyle{firststyle}{
  \fancyfoot[R]{}
  \fancyfoot[L]{}
}
\makeatother
\title{\textcolor{black}{Time-dependent  heterogeneity  leads to transient suppression of the COVID-19 epidemic, not herd immunity}}

\author[1,*]{Alexei V. Tkachenko}
\author[2,4,5,*]{Sergei Maslov}
\author[3]{Ahmed Elbanna}
\author[2]{George N.~Wong}
\author[2]{Zachary J.~Weiner}
\author[2,5] {Nigel Goldenfeld}
\affil[1]{Center for Functional Nanomaterials, Brookhaven National Laboratory, Upton, NY 11973, USA}
\affil[2]{Department of Physics, University of Illinois at Urbana-Champaign, Urbana, IL 61801, USA}
\affil[3]{Department of Civil Engineering, University of Illinois at Urbana-Champaign, Urbana, IL 61801, USA}
\affil[4]{Department of Bioengineering, University of Illinois at Urbana-Champaign, Urbana, IL 61801, USA}
\affil[5]{Carl R. Woese Institute for Genomic Biology, University of Illinois at Urbana-Champaign, Urbana, IL 61801, USA}

\leadauthor{Tkachenko}

\significancestatement{Epidemics generally spread through a succession
of waves that reflect factors on multiple time-scales.   Here, we develop a general
approach to encompass super-spreading and population heterogeneity, and demonstrate that a fragile state of
transient collective immunity (TCI) emerges well below the HIT during
early, high-paced stages of the epidemic. However, this is not an indication of herd immunity:
subsequent waves can and will emerge due to behavioral changes in the
population, driven (e.g.) by seasonal factors. Analysis of empirical data suggests that
even in locations with strong first waves of COVID-19, subsequent waves
will still emerge.}

\correspondingauthor{\textsuperscript{*}To whom correspondence should be addressed. E-mail: oleksiyt@bnl.gov or  maslov@illinois.edu}


\begin{abstract}

\textcolor{black} {Epidemics generally spread through a succession of
waves that reflect factors on multiple time-scales.  On short
time-scales, super-spreading events lead to burstiness and
overdispersion, while long-term persistent heterogeneity in
susceptibility is expected to lead to a reduction in the infection peak
and the herd immunity threshold (HIT). Here, we develop a general
approach to encompass both time-scales, including time variations in
individual social activity, and demonstrate how to incorporate them
phenomenologically into a wide class of epidemiological models through
parameterization.  We derive a non-linear dependence of the effective
reproduction number $R_e$ on the susceptible population fraction $S$.
We show that a state of transient collective immunity (TCI) emerges
well below the HIT during early, high-paced stages of the epidemic.
However, this is a fragile state that wanes over time due to changing
levels of social activity, and so the infection peak is not an
indication of herd immunity: subsequent waves can and will emerge due
to behavioral changes in the population, driven (e.g.) by seasonal
factors. Transient and long-term levels of heterogeneity are estimated
by using empirical data from the COVID-19 epidemic as well as from
real-life face-to-face contact networks. These results suggest that the
hardest-hit areas, such as NYC, have achieved TCI following the first
wave of the epidemic, but likely remain below the long-term HIT. Thus,
in contrast to some previous claims, these reqions can still experience
subsequent waves. }

\end{abstract}

\doi{\url{www.pnas.org/cgi/doi/10.1073/pnas.XXXXXXXXXX}}

\begin{document}

\maketitle

\thispagestyle{firststyle}
\ifthenelse{\boolean{shortarticle}}{\ifthenelse{\boolean{singlecolumn}}{\abscontentformatted}{\abscontent}}{}
The COVID-19 pandemic is nearly unprecedented in the level of
disruption it has caused globally, but also, potentially,  in the
degree to which it will change our understanding of epidemic dynamics
and the efficacy of various  mitigation strategies. Ever since the
pioneering works of Kermack and McKendrick
\cite{kermack1927contribution}, epidemiological models have been widely
and successfully used to quantify and predict the progression of
infectious diseases \cite{keeling2011modeling, rock2014dynamics,
MA2020129,Fraser2007,Chowell2017}. More recently, the important role in
epidemic spreading played by population heterogeneity and the complex
structure of social networks has been appreciated and highlighted in
multiple studies
\cite{Lloyd_May_Science_2001,may2001infection,Newman2002,Moreno2002,dezsHo2002halting,LloydSmith2005,Meyers_Newman_SARS_2005,Meyers_Newman_SARS_2005,
LloydSmith2005,Keeling_networks_2005,Meyer_fratility_2006,small2006super,empirical_Roy_Pascual,empirical_Stroud,bansal2007individual,Katriel_FSE_hetro,miller2012note,Meyers_future_epid_2012,pastor2015epidemic,gou2017heterogeneous,kim2018agent}.
However, integration of this conceptual progress into reliable,
predictive epidemiological models remains a formidable task. Among the
key effects of heterogeneity and social network structure are (i) the
role played by superspreaders and superspreading events during initial
outbreaks
\cite{LloydSmith2005,May_superspr_2005,SARS_superspr,Meyers_Newman_SARS_2005,small2006super,Super_Kucharski}
and (ii) a  substantial reduction of the final size of epidemic (FSE)
as well as the herd immunity threshold (HIT)
\cite{Newman2002,pastor2015epidemic,Moreno2002,bansal2007individual,Novozhilov,Katriel_FSE_hetro,miller2012note}.
The  COVID-19 pandemic has re-ignited interest in the effects of
heterogeneity of individual susceptibility to the disease, in
particular to the  possibility that it might lower both  HIT and  FSE
\cite{Beyond_R0,Herd_Gomes,Herd-Brennan,Ball_hetero_FSE,Herd_Science_2020}.
\textcolor{black}{In studying epidemics in heterogeneous populations  it
is important to emphasize the qualitative nature of two important
time-scales. First, overdispersion reflects short-term patterns of
behavior and even accidental events, and not the degree of persistent
population behavioral heterogeneity. Second, short-term overdispersion
is generally supposed to have a limited impact on the long term
epidemic dynamics, being important primarily in early outbreaks which
are dominated by super-spreading events. In other words, epidemics are
generally assumed to have distinct descriptions on widely-differing
time-scales.  In this paper, we attempt to provide a multi-scale theory
for epidemic progression and show that both overdispersion and
persistent heterogeneity affect the overall progression of the COVID-19
epidemic.  The significance of this multi-scale perspective is that it
provides a natural formalism to predict the occurrence and nature of
successive epidemic waves, even when it might seem that a first wave
has attained a state which could be mistaken for herd immunity.}

There are several existing approaches to model the effects of
heterogeneity on epidemic dynamics, each focusing on a different
characteristic and parameterization. In the first approach, one
stratifies the population into several demographic groups (e.g. by age),
and accounts for variation in susceptibility of these groups and their
mutual contact probabilities \cite{keeling2011modeling}. While this
approach represents many aspects of population dynamics beyond the
homogeneous and well-mixed assumption, it clearly does not encompass
the whole complexity of individual heterogeneity, interpersonal
communications and spatial and social structures. These details can be
addressed in a second approach, where one analyzes the epidemic
dynamics on real-world or artificial social networks
\cite{may2001infection,pastor2015epidemic,inferred_hetro_2013,Meyers_Newman_SARS_2005,Portland_net}.
Through elegant mathematics, it is possible to obtain detailed results
in idealized cases, including the mapping onto well-understood models
of statistical physics such as percolation \cite{Newman2002}.
\textcolor{black}{As demonstrated in Refs. \cite{miller2012note} the FSE
is a very robust property of the epidemic, insensitive to fine details
of its dynamics but defined by (i) distribution of susceptibilities in
the population
\cite{Katriel_FSE_hetro,hickson2014population,Novozhilov}; (ii)
correlations between infectivity and susceptibility. Importantly, it
does not depend on the part of heterogeneous infectivity that is not
correlated with susceptibility.
However, these approaches have so far been mostly limited to the
analysis of the FSE and distribution of outbreak sizes on a static
social network.}

For practical purposes, it is desirable to predict the complete
time-dependent dynamics of an epidemic, preferably by explicitly
including heterogeneity into classical well-mixed mean-field
compartmentalized models. \textcolor{black}{This approach was developed
some time ago in the context of epidemics on networks
\cite{Moreno2002,pastor2015epidemic} and the resulting mean-field
theory effectively reproduces the structure of heterogeneous well-mixed
models extensively studied in the applied mathematics literature
\cite{bansal2007individual,Novozhilov,Meyers_future_epid_2012,miller2012note,Katriel_FSE_hetro,gou2017heterogeneous}.
The overall impact of heterogeneity occurs because as the disease
spreads, it preferentially "vaccinates" the more susceptible
individuals, so the remaining population is less susceptible, and
spread is inhibited. This effect is further enhanced by a positive
correlation between infectivity and susceptibility. In the context of
static network models this correlation is perfect since both factors
are proportional to the degree of individual nodes. Ref.
\cite{gou2017heterogeneous} studied a hybrid model in which social
heterogeneity represented by network degree was combined with a
biological one.
These approaches have been recently applied in
the context of COVID-19 \cite{Herd_Science_2020,Herd_Gomes,Beyond_R0, weitz2020heterogeneity,neipel2020power}. Here, the conclusion was that the herd immunity threshold may be well below that expected in classical homogeneous
models.}

These approaches to heterogeneity delineate end-members of a continuum
of theories: overdispersion describing short-term, bursty dynamics
(e.g. due to super-spreader accidents), as opposed to {\em persistent
heterogeneity}, which is a long-term characteristic of an individual
and reflects behavioral propensity to (e.g.) socialize in large
gatherings without prudent social distancing.  Overdispersion is
usually modeled in terms of a negative binomial branching process
\cite{LloydSmith2005,May_superspr_2005,
SARS_superspr,Meyers_Newman_SARS_2005,small2006super,Super_Kucharski},
and is expected to be a much stronger source of variation compared to
the longer-term characteristics that reflect persistent heterogeneity.
\textcolor{black}{It is generally presumed that this short-term
overdispersion has no effect on the epidemic dynamics outside the early
outbreaks. Indeed, large variations in an individual's infectivity  would
average out as long as they are not correlated with susceptibility. But
since the initial exposure and  secondary infections are  separated by
a single  generation interval (typically about 5 days for COVID-19),
the levels of  individual social activity at those times are expected
to be correlated, and (at least partially) reflect short-term
overdispersion. How, then, can one  understand the epidemic progression
across various timescales, from  early stages of a fast exponential
growth to the final state of the herd immunity?}

Below, we present a comprehensive yet simple theory that accounts for
both social and biological aspects of heterogeneity, and predicts how
together they  modify early,  and intermediate  epidemic dynamics, as
well as global characteristics of the epidemic such as its HIT.
\textcolor{black}{Importantly, early epidemic dynamics may be sensitive
both to persistent heterogeneity and short-term overdispersion. As a
result, the apparent early-stage heterogeneity is typically enhanced
compared to its long-term persistent level. This may lead to a
suppression of the first wave of the epidemic due to reaching a novel
state that we call Transient Collective Immunity (TCI) determined by a
combination of short-term and long-term heterogeneity, whose threshold
is lower than the eventual HIT.  The implication is that the first
wave of an epidemic ends due to a combination of both persistent
heterogeneity and whatever mitigation constraints are imposed on the
population.  As the latter are relaxed by authorities or through
behavioral changes associated with seasonal factors, subsequent waves
can still occur.  Thus, TCI is dramatically different from the idea of
herd immunity due to heterogeneity.}

Our starting point is a generalized version of the heterogeneous
well-mixed theory  in the spirit of Ref. \cite{Moreno2002}, but we use
the age-of-infection approach \cite{kermack1927contribution} rather
than compartmentalized SIR/SEIR models of epidemic dynamics (see, e.g.
\cite{keeling2011modeling}). \textcolor{black}{Similar to multiple
previous studies, we first completely ignore any time dependence  of
individual susceptibilities and infectivities, focusing only on their
long-term   persistent components. This approach implicitly assumes
that any short-term overdispersion (responsible, e.g. for the
super-spreading phenomenon) is effectively averaged out. This is a
valid assumption if the large short-term variations in individual
infectivity are completely uncorrelated with an individual's susceptibility.
However, this approximation is hard to justify in the case of COVID-19.
Indeed, if the two are correlated on the timescale of a single
generation interval (5 days),    this will strongly affect the overall
epidemic dynamics. Therefore, our initial analysis  is eventually
expanded to a more general theory accounting for the non-negligible
effects of short-term overdispersion. In the case of persistent
heterogeneity we demonstrate how the} model can be recast into an
effective homogeneous  theory that can readily encompass a wide class
of epidemiological models, including various versions of the popular
SIR/SEIR approaches. Specific innovations that emerge from our analysis
are the non-linear dependence of the effective reproduction number
$R_e$ on the overall population fraction $S$ of susceptible
individuals, and another non-linear function $S_e$ that gives an
effective susceptible fraction, taking into account preferential
removal of highly susceptible individuals.

A convenient and practically useful aspect of this approach is that it
does  not require extensive additional calibration in order to be
applied to real data.  In the effort to make quantitative predictions
from epidemic models, accurate calibration is arguably the most
difficult step, but is necessary due to the extreme instability of
epidemic dynamics in both growth and decay phases
\cite{Wong2020,castro2020predictability}.  We find that with our
approach, the entire effect of heterogeneity is in many cases
well-characterized by a single parameter which we call  the {\it
immunity factor}  $\lambda$. It is related to the statistical properties of
heterogeneous susceptibility across the population and to its
correlation with individual infectivity. The immunity factor
determines the rate at which $R_e$ drops during the early stages of the
epidemic as the pool of susceptibles is being depleted:  $R_e\approx
R_0(1-\lambda (1-S))$. Beyond this early linear regime, for an
important   case of gamma-distributed individual susceptibilities, we
show that the classical proportionality, $R_e=R_0 S$, transforms into a
power-law scaling relationship  $R_e=R_0S^\lambda$. This leads to a
modified version of the result for the herd immunity threshold, \textcolor{black}{
$1-S_0=1-R_0^{-1/\lambda}$}.

Heterogeneity in  the susceptibility of individual members of the
population has several different contributions: (i) biological, which
takes into account differences in factors such as strength of immune
response, genetics, age, and comorbidities; and  (ii) social,
reflecting differences in the number \textcolor{black}{and frequency} of close contacts of different people.
The immunity factor $\lambda$ in our model combines these sources of
heterogeneous susceptibility as well as its correlation with individual
infectivity. As we demonstrate, under certain assumptions the immunity
factor is simply a product of social and biological contributions:
$\lambda=\lambda_s\lambda_b $. In our study, we leverage existing
studies of real-life face-to-face contact networks
\cite{Crowd,Sneppen2020,Barrat_proximity_2017,UK_social_Keeling,
Portland_net,Meyers_Newman_SARS_2005,bansal2007individual} to estimate
the social contribution to heterogeneous susceptibility, and the
corresponding  immunity factor $\lambda_s$. The biological
contribution, $\lambda_b$, is expected to depend on specific details of
each infection. For the case of COVID-19, \textcolor{black}{there is little indication that biological variations in susceptibility, unrelated to one's social activity, play a significant role in the epidemic dynamics.}

To test this theory, we use empirical data for the COVID-19 epidemic to
independently estimate the immunity factor $\lambda$. In particular, we
apply our previously-described epidemic model that features
multi-channel Bayesian calibration \cite{Wong2020} to describe epidemic
dynamics in New York City and Chicago. This model uses high quality
data on hospitalizations,  Intensive Care Unit (ICU) occupancy and
daily deaths to extract the underlying $R_e(S)$ dependence in each of
two cities. In addition, we perform a similar analysis of data on
individual states in the USA, using data generated by the model in Ref.
\cite{Juliette2020report}. Using both approaches,  we find that the
locations  that were severely impacted by the COVID-19 epidemic show a
more pronounced  reduction of the effective reproduction number. This
effect is much stronger than predicted by classical homogeneous models,
suggesting a significant role  of heterogeneity. The estimated immunity
factor ranges between  $4$ and $5$. \textcolor{black}{Importantly, this
represents a transient value of the parameter $\lambda$ observed on
intermediate timescales and dependent on both persistent and short-term
heterogeneity. Our estimates of the long-term, value of the immunity
factor, defined by persistent heterogeneity only, is considerably
lower: $\lambda_{\infty} \simeq 2$. This difference explain why
achieving the state of transient collective immunity after the first
wave of the epidemic does not imply long-term herd immunity.}
%

Finally, we integrate the persistent heterogeneity theory into our
time-of-infection epidemiological model \cite{Wong2020}, and project
possible outcomes of the second wave of the COVID-19 epidemic in
\textcolor{black}{during the summer months} in NYC and Chicago, using
data up to the end of May 2020. By considering the worst-case scenario
of a full relaxation of any currently imposed mitigation, we find that
the results of the heterogeneity-modified  model significantly modify
the results from the homogeneous mode.  In particular,  based on our
estimate of the immunity factor, we expect virtually no second wave in
NYC \textcolor{black}{in the immediate future}, indicating that the
\textcolor{black}{TCI} has likely been achieved there. Chicago, on the
other hand, has not passed  the \textcolor{black}{TCI} threshold that we
infer, but the effects of heterogeneity would still result in a
substantial reduction of the magnitude of the second wave there, even
under the worst-case scenario. This, in turn, suggests that the second
wave can be completely eliminated in such medium-hit locations, if
appropriate and economically mild mitigation measures are adopted,
including e.g. mask wearing, contact tracing, and targeted limitation
of potential super-spreading events, through limitations on indoor
bars, dining and other venues. \textcolor{black}{We further investigate
the issue of fragility of collective immunity in heterogeneous
populations. By allowing rewiring of the social network with time, we
demonstrate that the TCI may wane, much like an individual's acquired
immunity may wane due to biological factors. This phenomenon would
result in the emergence of secondary epidemic waves after a substantial
period of low infection counts.}

\section*{Theory of epidemics in populations \textcolor{black}{with persistent  heterogeneity}}
Following in the footsteps of
Refs.\cite{Moreno2002,pastor2015epidemic,bansal2007individual,Novozhilov,Meyers_future_epid_2012,Herd_Gomes,gou2017heterogeneous},
we consider the spread of an epidemic in a population of individuals
who exhibit significant \textcolor{black}{persistent} heterogeneity in their susceptibilities to
infection $\alpha$. This heterogeneity may be
biological or social in origin, and we assume these factors are independent: $\alpha=\alpha_b\alpha_s$.
\textcolor{black}{Effects of possible correlations between $\alpha_b$ and $\alpha_s$ have been discussed in Ref. \cite{gou2017heterogeneous}.}
The biologically-driven heterogeneous susceptibility $\alpha_b$ is shaped
by variations of several intrinsic factors such as the strength of
individuals' immune responses, age, or genetics. In contrast, the
socially-driven heterogeneous susceptibility $\alpha_s$ is shaped by
extrinsic factors, such as differences in individuals' social
interaction patterns (their degree in the network of social
interactions).
Furthermore, individuals' different risk perceptions and attitudes
towards social distancing may further amplify variations in
socially-driven susceptibility heterogeneity.
We only focus on susceptibility that is a persistent property of an
individual. For example, people who have elevated occupational
hazards, such as healthcare workers, typically have higher, steady
values of $\alpha_s$.  Similarly, people with low immune response,
highly social individuals (hubs in social networks), or scofflaws would
all be characterized by above-average overall susceptibility $\alpha$.

In this work, we group individuals into sub-populations with similar
values of $\alpha$ and describe the heterogeneity of the overall
population by the probability density function (pdf) of this parameter,
$f(\alpha)$. Since $\alpha$ is a relative measure of individual
susceptibilities, without loss of generality we set $\langle \alpha
\rangle\equiv\int_0^{\infty}\alpha f(\alpha)d\alpha=1$. Each person is also
assigned an individual reproduction number $R_i$, which is an
expected number of people that this person would infect in a
fully susceptible population with $\langle \alpha \rangle=1$.
Accordingly, from  each sub-population with susceptibility $\alpha$
there is a respective  mean \textcolor{black}{infectivity or}  reproductive number $R_\alpha$. Any
correlations between individual susceptibility and infectivity will
significantly impact the epidemic dynamics.
Such correlations are an integral part of most network-based
epidemiological models  due to the assumed reciprocity in underlying
social interactions, which leads to  $R_{\alpha} \sim \alpha$
\cite{Newman2002,Moreno2002,pastor2015epidemic}. In reality, not all transmissions
involve face-to-face contacts, and biological susceptibility need not
be strongly correlated with infectivity. Therefore, it is reasonable to
expect only a partial correlation between $\alpha$ and $R_\alpha$.

Let $S_\alpha(t)$ be the fraction of susceptible individuals in the
subpopulation with susceptibility $\alpha$, and let
$j_\alpha(t)=-\dot{S}_\alpha$ be the corresponding daily incidence rate,
i.e., the fraction of newly infected individuals per day in that
sub-population.  At the start of the epidemic, we assume everyone is
susceptible to infection: $S_\alpha(0)=1$.  The course of the epidemic
is described by the following age-of-infection model:
 \begin{equation}
\label{generic}
    -\frac {dS_\alpha}{dt}=j_\alpha(t)=\alpha S_\alpha(t) J(t)
\end{equation}
\begin{equation}
\label{I}
    J(t)=  \int_{0}^{\infty} \left \langle R_\alpha K(\tau)j_\alpha(t-\tau)\right \rangle  d\tau
\end{equation}
Here $t$ is the physical time and $\tau$ is the time since infection
for an individual. $\langle \ldots \rangle$ represents averaging
over $\alpha$ with pdf $f(\alpha)$.  $J(t)$ represents the mean daily
attack rate, i.e. \textcolor{black}{hypothetical incidence rate in a fully susceptible homogeneous population with  $\alpha=1$}. $K(\tau)$ is the
distribution of the generation interval, which we assume to be  independent of
$\alpha$ for the sake of simplicity.

According to Eq.~(\ref{generic}), the susceptible subpopulation for any $\alpha$ can be expressed as
\begin{equation}
S_\alpha(t)=\exp(-\alpha Z(t)) .
\label{s_alpha_vs_t}
\end{equation}
Here $Z(t)\equiv\int_0^t J(t') dt'$.
The total susceptible fraction of the population is related to the
moment generating function $M_\alpha$ of the distribution $f(\alpha)$
(i.e., the Laplace transform of $f(\alpha)$) according to:
\begin{equation}
\label{S_def}
    S(t)= \int_0^\infty f(\alpha) e^{-\alpha Z(t)} d\alpha=M_\alpha(-Z(t))
\end{equation}

Similarly, the  effective reproductive number $R_e$ can be expressed in
terms of the parameter $Z$:
\begin{equation}
\label{Re}
    R_e(t)=\int_0^\infty \alpha R_\alpha f(\alpha) e^{-\alpha Z(t)} d\alpha
\end{equation}
Note that for $Z=0$, this expression gives the basic  reproduction
number  $R_0=\langle\alpha R_\alpha\rangle$. Since both $S$ and $R_e$
depend on time only through $Z(t)$, Eqs. (\ref{S_def})--(\ref{Re})
establish a parametric relationship between these two important
quantities during the time course of an epidemic. In contrast to the
classical case when these two quantities are simply proportional to
each other, i.e. $R_e=S R_0$,  the relationship in the present theory
is non-linear due to heterogeneity. Now one  can re-write the renewal
equation for the force of infection  in the same form as if this were  a homogeneous problem:
\begin{equation}
\label{attack0}
    J(t)=\int_{0}^{\infty} d\tau K(\tau)R_e(t-\tau)J(t-\tau)
\end{equation}
Furthermore, by averaging Eq. (1) over all values of $\alpha$ one arrives at the following heterogeneity-induced
modification to the relationship between the force of infection  and  incidence
rate:
\begin{equation}
\label{incident}
    \frac{ dS}{d t}=- S_e J
\end{equation}
Here
\begin{equation}
\label{eq:Se}
    S_e(t)=\int_0^\infty \alpha f(\alpha) e^{-\alpha Z(t)} d\alpha=-\frac{dM_\alpha(-Z(t))}{dZ}
\end{equation}
is the effective susceptible fraction of the population, which is less
than $S$ due to the disproportionate removal of highly susceptible
individuals. Just as with $R_e$, it is a non-linear function of $S$,
defined parametrically by Eqs. (\ref{S_def},\ref{eq:Se}). Further
generalization of this    theory  for the time-modulated
age-of-infection model is presented in the {\it Supporting Information Appendix (SI Appendix)}.
There, we also discuss the adaptation of this approach for the
important special case of a compartmentalized SIR/SEIR model.  Such
non-linear modifications to homogeneous epidemiological  models have
been proposed  in the past, both as  plausible   descriptions of
heterogeneous populations and in other contexts
\cite{bansal2007individual,empirical_Stroud, empirical_Roy_Pascual}.
However, those empirical models exhibited  a limited range of
applicability \cite{bansal2007individual} and   have not had a solid
mechanistic foundation, with  a noticeable exception of  a special case
of SIR model without correlation between susceptibility and infectivity
studied in  Ref. \cite{Novozhilov}. Our approach is more general: it
provides the exact mapping of a wide class of heterogeneous well-mixed
models onto  homogeneous ones, and provides a specific relationship
between the underlying statistics of $\alpha$ and $R_\alpha$ and the
non-linear functions $R_e(S)$ and $S_e(S)$.

One of the striking consequences of the non-linearity of $R_e(S)$  is
that  the effective reproduction number could be  decreasing at the
early stages of an epidemic significantly faster than predicted by
homogeneous models. Specifically, for $1-S(t) \simeq Z(t) \ll 1$ one
can linearize the effective reproduction number as
\begin{equation}
\label{lambda1}
R_e\approx R_0 (1-\lambda (1-S))
\end{equation}
We named the coefficient $\lambda$ the {\it immunity factor} because it
quantifies the effect that
a gradual build-up of   population immunity
has on the spread of an epidemic. The classical value
of $\lambda$ is 1, but it may be significantly larger in a heterogeneous
case.
\textcolor{black}{By linearizing  Eq. (\ref{Re}) in terms of $1-S \simeq
Z  \ll 1$ and dividing the result by $R_0=\langle \alpha R_{\alpha}
\rangle$ one gets:}
 \begin{equation}
\label{lambda2}
\lambda= \frac{\langle \alpha^2 R_\alpha \rangle}{\langle \alpha R_\alpha\rangle}
\end{equation}
As one can see, the value of the immunity factor, thus depends  both on
the statistics of susceptibility $\alpha$, and on its correlation with
infectivity $R_{\alpha}$.

We previously defined the  overall susceptibility as a combination of
biological and social factors: $\alpha=  \alpha_s \alpha_b$ Here
$\alpha_s$ is a measure of the overall social connectivity or activity
of an individual, such as the cumulative time of close contact with
other individuals averaged over a sufficiently long time interval
(known as node strength in network science). Since the contribution of interpersonal
contacts to an epidemic spread is mostly reciprocal, we
assume $R_\alpha \sim \alpha_s$. On the other hand, in our analysis we
neglect a correlation between biological susceptibility and
infectivity, as well as between $\alpha_b$ and  $\alpha_s$. Under these
approximations,  the immunity factor itself is a product of biological
and social  contributions,  $\lambda=\lambda_b \lambda_s$. Each of them
can be expressed in terms of leading  moments of  $\alpha_b$ and
$\alpha_s$, respectively:
\begin{equation}
\label{lambda_b}
    {\lambda}_{b}  =\frac{\langle \alpha_{b}^2\rangle}{\langle \alpha_{b}\rangle^2}=1+CV_b^2
\end{equation}
\begin{equation}
\label{lambda_s}
    \lambda_s=\frac{\langle  \alpha_{s}^3\rangle}{\langle  \alpha_{s}\rangle\langle  \alpha_{s}^2\rangle}= 1+\frac{CV_s^2(2+\gamma_s CV_s)}{1+CV_s^2}
\end{equation}
\textcolor{black}{These equations follow from  Eq. \ref{lambda2} in the
limit $R_{\alpha} =\mathrm{const}$ and $R_{\alpha} \sim \alpha$
respectively.} \textcolor{black}{Although these equations resemble
classical results for $R_0$ in heterogeneous networks
\cite{may2001infection,Moreno2002,Newman2002,pastor2015epidemic}, here
they describe a completely different effect of suppression of $R_e$ in
response to depletion of susceptible population $S$. That is why
$\lambda_s$ in Eq. \ref{lambda_s} is proportional to the third moment
of $\alpha_s$ instead of the second moment in the case of $R_0=\langle
\alpha R_{\alpha} \rangle \sim \alpha_s^2$.} Note that the biological
contribution to the immunity factor depends only on the coefficient of
variation  $CV_b$ of  $\alpha_b$. On the other hand, the social factor
$\lambda_s$ depends both on the coefficient of variation $CV_s$ and the
skewness $\gamma_s$ of the distribution of $\alpha_s$. Due to our
normalization, $\langle \alpha_s \rangle \langle\alpha_b \rangle\approx
\langle \alpha_s \alpha_b \rangle= \langle \alpha \rangle =1$.

The  relative importance  of biological and social contributions to the
overall heterogeneity of  $\alpha$ may be characterized  by a single
parameter $\chi$. For a log-normal distribution  of $\alpha_b$, $\chi$
appears as a scaling exponent between infectivity and susceptibility:
$R_\alpha \sim \alpha^\chi$ (see {\it SI Appendix} for details). The
corresponding expression for the  overall immunity factor is
$\lambda=\langle\alpha^{2+\chi} \rangle / \langle\alpha^{1+\chi}
\rangle $. The limit  $\chi=0$ corresponds to a predominantly
biological source of heterogeneity, i.e., $\lambda \approx \lambda_b =
1+CV^2_\alpha$, where $CV_{\alpha}$ is the coefficient of variation for
the overall susceptibility. In the opposite limit $\chi=1$, population
heterogeneity is primarily of social origin, hence  $\lambda \approx
\lambda_s$ is affected by both $CV_\alpha$ and the skewness
$\gamma_{\alpha}$ of the pdf $f(\alpha)$. The biological contribution
$\lambda_{b}$ depends on specific biological details of the disease and
thus is unlikely to be as universal and robust as the social one. For
the COVID-19 epidemic, there is no strong evidence of a wide variation
in attack rates unrelated to social activity, geographic location,  or
socioeconomic status. For instance,  there is very little  age
variability in COVID-19 prevalence as  reported by the NYC Department
of Health \cite{Nychealth} based on the serological survey that
followed the first wave of the epidemic. Therefore, below
we will largely ignore possible  biological heterogeneity, and focus on
social heterogeneity.


 \begin{figure}[ht!]
\centering
\includegraphics[width=1\columnwidth]{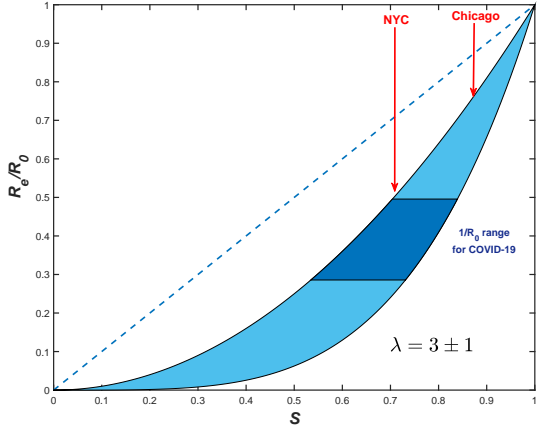}
\caption{ $R_e/R$ vs $S$ dependence for gamma-distributed
susceptibility with  \textcolor{black}{$\lambda=3\pm 1$} (blue area). The dashed line shows
the classical homogeneous result, $R_e=R_0S$. Note a substantial
reduction of $R_e$ for COVID-19 epidemic expected in both NYC and Chicago, compared to
that value. Approximate fractions of susceptible populations,  $S$,
for both cities are estimated as of the end of   May 2020,  by using
the   model described in  Ref.  \cite{Wong2020}.   }
\label{fig:RvsS}
\end{figure}

So far, our discussion has focused on the early stages of epidemics,
when the $R_e(S)$ dependence is given by a linearized expression
Eq.~(\ref{lambda1}).  To describe the non-linear regime, we consider a
gamma-distributed susceptibility: $f(\alpha)\sim
\alpha^{1/\eta-1}\exp(-\alpha/\eta)$, where  $\eta= CV_\alpha^2$. In
this case, according to Eqs.~(\ref{S_def}) and (\ref{Re}), $R_e$, $S_e$
and $S$ are related by scaling relationships \textcolor{black}{(see {\it SI Appendix})}:
\begin{equation}
\label{scaling0}
    S_e(S)=S^{1+\eta}
\end{equation}
\noindent
and
\begin{equation}
    R_e(S)=R_0S^\lambda
\label{scaling}
\end{equation}
The exponent $\lambda= 1+(1+\chi)CV_\alpha^2=1+(1+\chi)\eta $ coincides
with the  early-epidemics immunity factor  defined in
Eqs.~(\ref{lambda1})--(\ref{lambda2}) for a general case.  Note that
without correlation ($\chi=0$), both scaling exponents would be the
same; this result has been previously obtained for the SIR model in
Ref.~\cite{Novozhilov} \textcolor{black}{and more recently reproduced in
Ref. \cite{weitz2020heterogeneity} in the context of COVID-19.} The
scaling  behavior $R_e(S)$ is shown in Fig. \ref{fig:RvsS} for
$\lambda =3 \pm 1$ .
This function  is dramatically   different from the classical linear
dependence $R_e=SR_0$. To emphasize the importance of this difference,
we indicate   the  estimated  fractions of the population  in  New York
City and Chicago susceptible to COVID-19, as of the end of May  2020.

Eq. (\ref{scaling})  immediately leads to  a major revision of the
classical result for the herd immunity threshold
\textcolor{black}{$1-S_0=1-1/R_0$. $S_0$ is the fraction of susceptible
population at which the exponential growth stops, while $1-S_0$ is the
relative size of the epidemic at that time.}  By setting $R_e=1$ in Eq.
(\ref{scaling}), we obtain:
\begin{equation}
\label{eq:HIT}
    1-S_0=1-\left(\frac{1}{R_0} \right) ^{1/\lambda}
\end{equation}
\textcolor{black}{As we were finalizing this paper for public release, a
preprint by Aguas et al.  appeared online
\cite{Aguas_Gomes_Hetero_2020} that independently obtained our Eqs.
(\ref{scaling}, \ref{eq:HIT}) for gamma-distributed susceptibilities.
The same result has also been recently obtained in Ref.
\cite{neipel2020power}. Note however that the full quasi-homogeneous
description of the epidemic dynamics requires both $R_e(S)$ and
$S_e(S)$, that in the general case is characterized by  different scaling
exponents.
}


Our focus on the gamma  distribution is well justified by  the
observation that the social strength $\alpha_s$ is approximately
exponentially distributed, i.e., it is a specific case of the gamma
distribution with $\eta=CV_\alpha^2=1$ (see more discussion of this in
the next section). A moderate biological heterogeneity would lead to an
increase of the  overall $CV_\alpha$, but the pdf $f(\alpha)$ will
still be close to the gamma distribution family.  From the conceptual
point of view, it is nevertheless important to understand how the
function $R_e(S)$ would change if $f(\alpha)$ had a different
functional form. In {\it SI Appendix}, we present analytic and
numerical calculations for two other families of distributions: (i) an
exponentially bounded  power law $f(\alpha)\sim
e^{-\alpha/\alpha_+}/\alpha^q$  ($q \ge 1$, with an additional cut-off
at lower values of $\alpha$) and (ii) the log-normal distribution.  In
addition, we give an  approximate analytic result that generalizes
Eq.~(\ref{scaling}) for an arbitrary skewness of $f(\alpha)$. This
generalization works remarkably well for all three families of
distributions analyzed in this work. As suggested by Eqs.
(\ref{lambda_b}-{\ref{lambda_s}}), as the  distribution becomes more
skewed, the range  between the $\chi=0$ and $\chi=1$ curves broadens.
For instance, for distributions dominated by a power law, $f(\alpha)
\sim 1/\alpha^q$,  with $3 < q < 4$ and $\chi=1$, $\lambda$ diverges
even though  $CV_\alpha$ remains finite. This represents a crossover to
the   regime of so-called scale-free networks ($2 \le q \le 3$), which
are characterized by zero epidemic threshold yet strongly self-limited
dynamics: the epidemic effectively kills itself by  immunizing the hubs
on the network \cite{Pastor_Scale_free,Moreno2002,pastor2015epidemic}.

\section*{\textcolor{black}{Role of short-term variations in social activity}}

\textcolor{black} {Short-term overdispersion in transmission is commonly
presumed to have no effect on the overall epidemic dynamics, aside from
the early outbreak often dominated by superspreaders. This would indeed
be the case if overdispersed transmission were completely uncorrelated
with individual susceptibility. But since the timescale for an
individual's infectivity (about 2 days) is comparable to a single
generation interval (about 5 days)  for the COVID-19 epidemics,
ignoring such correlations appears unreasonable. We therefore developed
a generalization of the theory presented in the previous section, that
takes into account  a time dependence of individual susceptibilities
and infectivities, as well as temporal correlations between them. The
theory is presented using a path-integral formulation in {\it SI
Appendix}. Here we present several important results directly related
to the transient suppression of an epidemic and differentiate these
effects from herd immunity. }

\textcolor{black} { Since fast variations are primarily caused by bursty
dynamics of social interactions
\cite{barabasi2005origin,kossinets2006empirical,rybski2009scaling,saramaki2015seconds},
and since heterogeneous biological susceptibility appears subdominant
in the context of COVID-19, we set $\alpha_b=1$ for all individuals.
So  $\alpha$ has purely social origin. Let  $a_i(t)=\alpha_i+\delta
a_i(t)$ be the time-dependent  susceptibility of a person, which we
associate with variable level of social activity. Hence, the same
function determines also individual  infectivity $a(t)R$  around time
$t$. Interestingly,  even the  classical  result for basic reproduction
number  in a heterogeneous system, $R_0=R\langle \alpha^2 \rangle $,
needs to be modified due to correlated short-term variations in social
activity:
\begin{equation}
\label{R0}
    R_0=R\left(\langle \alpha^2\rangle+\overline{\delta a_i^2}\right)
\end{equation}}


\textcolor{black}{In the time-dependent generalization of our  theory
$R_e$ and $S$ no longer have a fixed functional relationship between
them. Instead, this relationship becomes non-local in time. For
instance, our   result for the suppression of $R_e$ at the early stages
of the epidemic is still formally valid, but the effective value of
immunity factor $\lambda$ becomes time dependent, and
Eqs.(\ref{lambda1})-(\ref{lambda2}) become
\begin{align}
\label{lambda_eff}
\lambda_{\rm eff}(t)&=\lambda_\infty+\frac{1}{1-S(t)}\int_{0}^{\infty} \delta\lambda(t,t')J(t-t')dt'  \\
  \label{lambda_inf}
  \lambda_\infty&=\frac{\langle \alpha^3\rangle +\overline{\alpha_i \delta a_i^2}}{\langle \alpha^2\rangle +\overline{ \delta a_i^2}}\\
\label{delta_lambda}
\delta\lambda(t,t')&=\frac{\overline{\delta a^2_i(t)\delta a_i(t-t')}}{\langle \alpha^2\rangle +\overline{ \delta a_i^2}}
\end{align}
Constant $\lambda_\infty$ reflects suppression of $R_e$ due to the
build-up of the  long-term collective immunity. On the other hand, the
time-dependent term  $\delta\lambda(t')$ leads to an additional
suppression of $R_e$ over intermediate timescales. This term has likely
played a significant role in shaping the transient self-limiting
dynamics during the first  wave of COVID-19 epidemic in some hard-hit
locations.}

\textcolor{black}{
Note that according to Eq. (\ref{lambda_eff}), $\delta \lambda(t')$ is
being averaged with the weight proportional to the force of infection
$J(t-t')$ since $1-S(t)\approx \int_0^{\infty} J(y-t')dt'$. Since
$\delta \lambda(t,t') $ decreases with time difference $t'$, its effect
on  $\lambda_{\rm eff}$ should be the strongest during the initial
period of fast exponential growth. The  initial suppression of the
epidemic is caused by  the combined effect of mitigation measures and
both terms in  $\lambda_{\rm eff}$.  Since $\lambda_{\rm eff}
>\lambda_\infty$, the population may reach the state of Transient
Collective Immunity (TCI) earlier than the actual long term Herd
Immunity determined by persistent heterogeneity. However, this state is
fragile and may wane with time.  Specifically,  as $J(t)$ drops after
the first wave,    the second term in Eq. (\ref{lambda_eff})  gradually
decays, bring $\lambda_{\rm eff}(t)$ closer to $\lambda_\infty$.
According to  Eq.(\ref{delta_lambda}), it is the correlation time of
bursty social activity $\delta a(t)$ that  sets the timescale over
which this TCI state  deteriorates, and the new epidemic wave may get
ignited.   The relationship between this relaxation time and the
duration of a single  epidemic wave also determines the typical value
of $\lambda_{\rm eff}$ during that wave.
}

\begin{figure}[ht!]
\centering
\includegraphics[width=1\columnwidth]{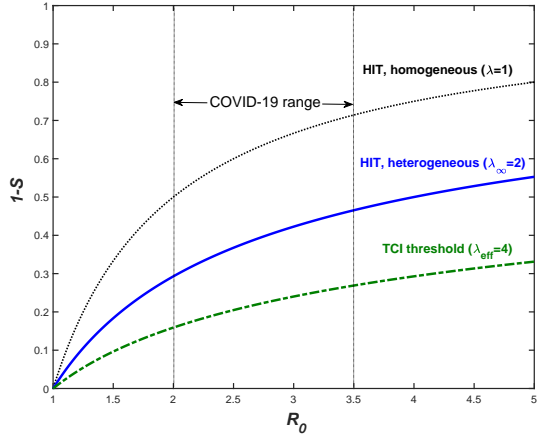}
\caption{ \textcolor{black}{ TCI threshold (dot-dashed), long-term heterogeneous HIT (solid), and homogeneous HIT (dotted)  for various values of $R_0$ (x-axis). HIT (solid line) is determined by  persistent heterogeneity. The corresponding immunity factor $\lambda_\infty\approx 2$ is estimated  from  empirical face-to-face contact network. For transient  behavior $\lambda_{\rm eff} \approx 4$ is assumed based on analysis of empirical data for COVID-19 epidemic in select locations. }}
\label{fig:HIT}
\end{figure}

\textcolor{black} { Despite a  large number of empirical studies of
contact networks
\cite{barabasi2005origin,kossinets2006empirical,rybski2009scaling,saramaki2015seconds},
information  about the temporal  correlations in $\alpha(t)$ or its
proxies  remains  limited.  On the other hand, much  more is known
about parameters of  persistent heterogeneity.   }Recently, real-world
networks of face-to-face communications have been studied using a
variety of tools, including RFID devices \cite{Crowd}, Bluetooth and
Wi-Fi wearable tags, smartphone apps
\cite{Sneppen2020,Barrat_proximity_2017}, as well as census data and
personal surveys \cite{UK_social_Keeling,
Portland_net,Meyers_Newman_SARS_2005}. Despite coming from a wide
variety of contexts, the major features of contact networks are
remarkably robust. In particular, pdfs of both the degree (the number
of contacts per person), and the node strength plotted in log-log
coordinates appear nearly constant followed by a sharp fall after a
certain upper cut-off. This behavior is generally consistent with an
exponential distribution in $f_s(\alpha_s)$ \cite{Sneppen2020,
bansal2007individual, UK_social_Keeling}, $f(\alpha)\sim
e^{-\alpha/\langle \alpha \rangle }$. \textcolor{black} {That sets the
value of $\eta=CV_\alpha^2\approx 1$. If not for short-term
overdispersion,  that would yield  $\lambda =\langle \alpha_s^3
\rangle/\langle \alpha_s^2 \rangle = 3!/2!=3$ according to Eq.
(\ref{lambda_s}). However, with temporal effects  taken into account,
the build-up of  long-term collective immunity is determined by
$\lambda_{\rm eff}(\infty)=\lambda_{\infty}$.  In order to estimate it,
we make a simple  model assumption that the  short-term overdispersion
for a particular individual is proportional to the  persistent value of
that person's   social activity:  $\overline{\delta a_i^2}\sim
\alpha_i$. This leads to \begin{equation}
  \lambda_\infty=1+\eta(1+\chi*)
\end{equation}
Here $\chi^*=\langle \alpha^2\rangle/\overline{ a_i(t)^2}$ is a
parameter that measures the relative strength of persistent
heterogeneity and the overdispersion on the timescale of a single
generation interval. Note that formally we recover our original result
for $\lambda$ in the purely persistent case, with $\chi^*$ replacing
the
parameter $\chi$ that originally  quantified the correlation between
infectivity and susceptibility. By assuming the limit of  strong
short-term overdispersion ($\chi^*\ll 1$), we obtain
$\lambda_\infty\approx 2$.  As shown in {\it SI Appendix}, the very
same value of $\lambda_\infty $ should be used as a scaling exponent
for long-term behavior  of $R_e(S)$. Therefore, HIT is set by
Eq.(\ref{eq:HIT}) with $\lambda=\lambda_\infty\approx 2$.  Its value is
plotted vs. $R_0 $ in Fig. \ref{fig:HIT}, along with the homogeneous
result, and the estimated threshold of $TCI$. To estimate the corresponding
transient  immunity factor $\lambda_{\rm eff}$, we analyze the
empirical data for the first wave of the COVID-19 epidemic below. }

\section*{Application to the  COVID-19 epidemic}
\begin{figure}[ht]
\centering
\includegraphics[width=0.95\columnwidth]{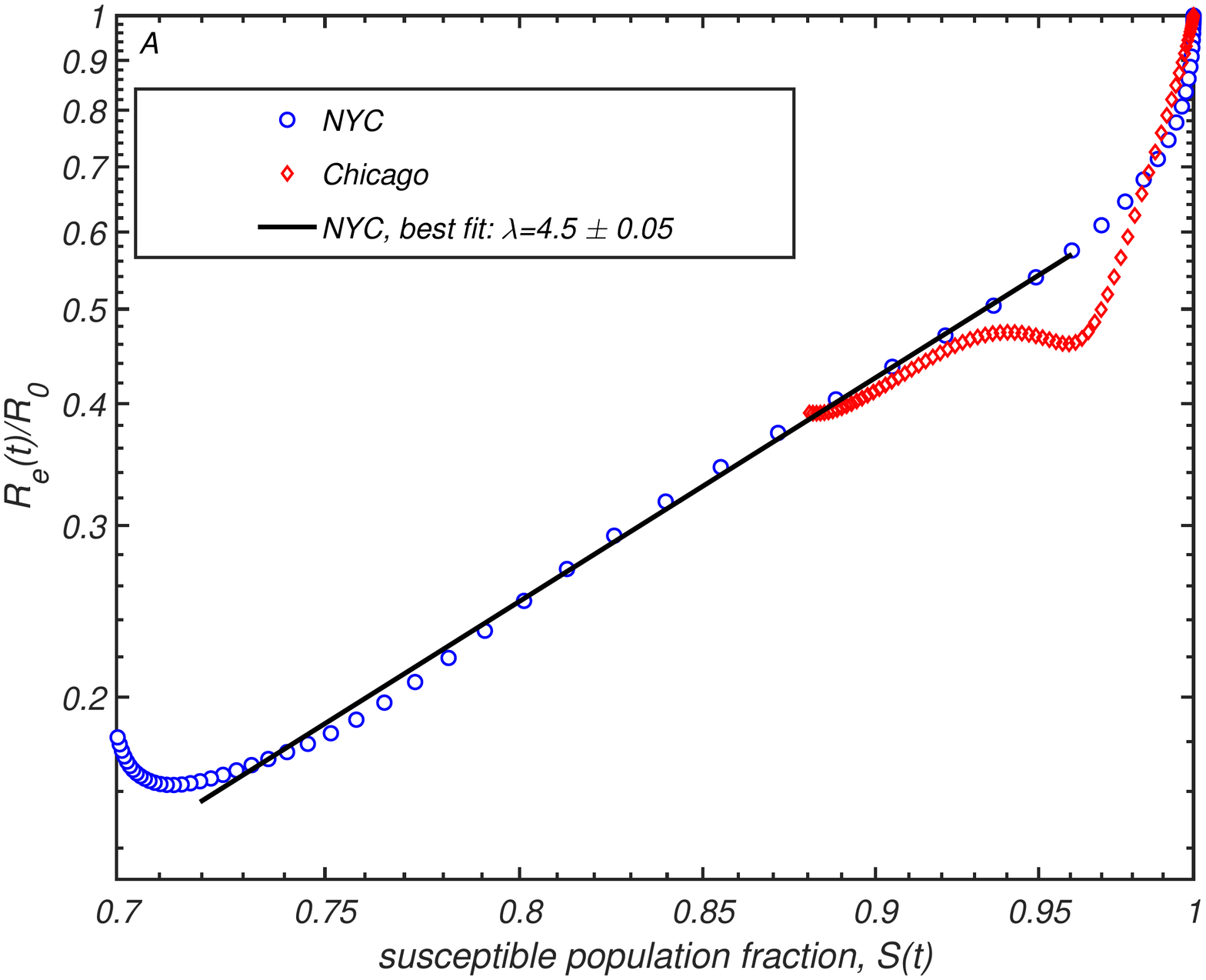}
\includegraphics[width=0.95\columnwidth]{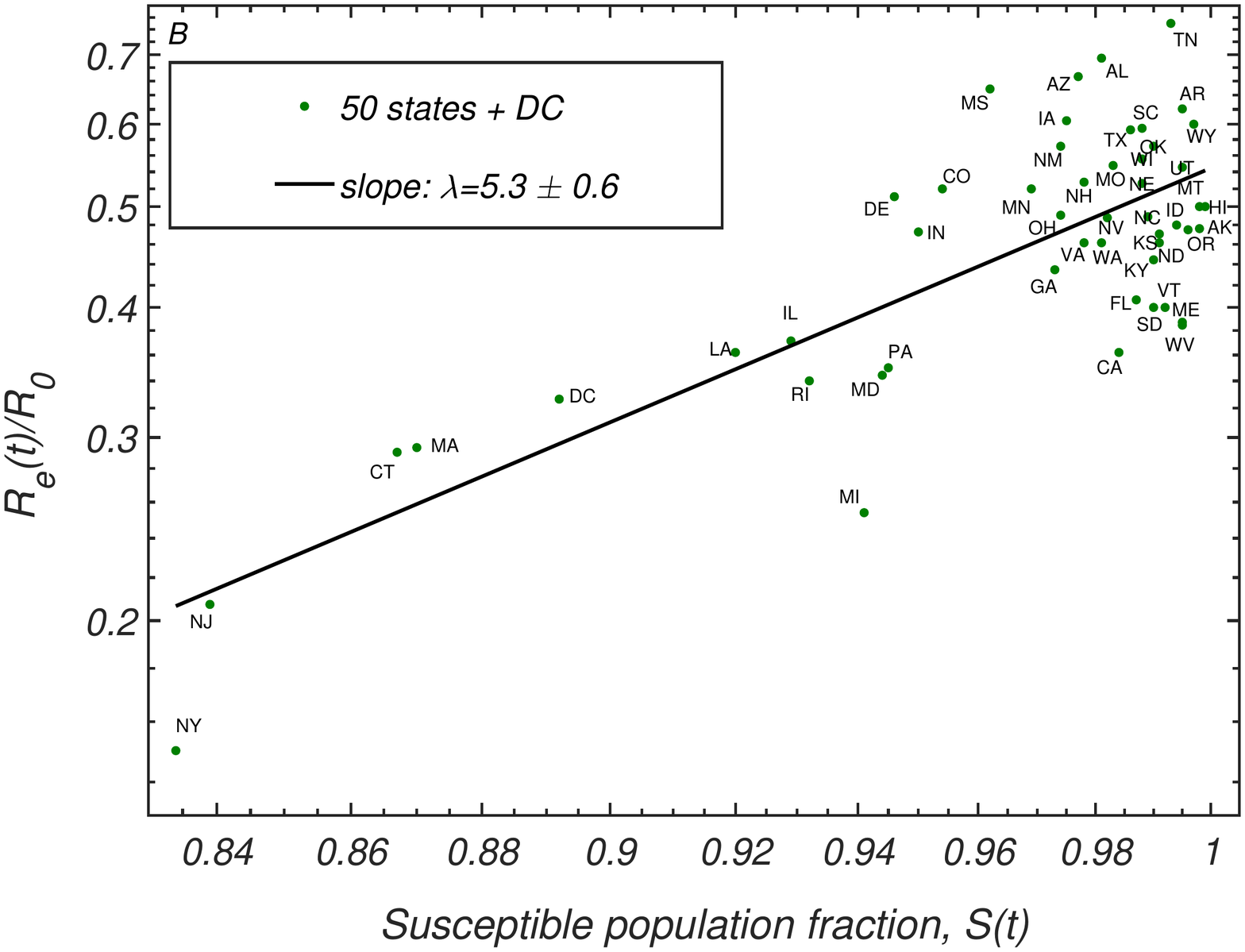}
\caption{Correlation between the relative reduction in the effective
reproduction number $R_e(t)/R_0$ (y-axis) with the susceptible
population $S(t)$. In Panel A, we present the progression of these two
quantities for New York City and Chicago, as given by the
epidemiological model described in Ref. \cite{Wong2020}. Panel B shows
the scatter plot of $R_e(t_0)/R_0$ and $S(t_0)$ in individual states of
the US, evaluated in Ref. \cite{Juliette2020report} ($t_0$ is the
latest date covered in that study).
}
\label{fig:COVID_Re_vs_S}
\end{figure}

The COVID-19 epidemic reached the US in early 2020, and by March it was
rapidly spreading across multiple states. The early dynamics was
characterized by a rapid rise in the number of cases with doubling
times as low as $2$ days.
In response to this, the majority of states imposed a broad range of
mitigation measures including school closures, limits on public
gatherings, and Stay-at-Home orders. In many regions, especially the
hardest-hit ones like New York City, people started to practice some
degree of social distancing even before government-mandated mitigation.
In order to quantify the effects of heterogeneity on the spread of the
COVID-19 epidemic, we  apply the Bayesian age-of-infection model
described in Ref. \cite{Wong2020} to  New York City and Chicago. For
both cities, we have access to reliable time series data on
hospitalization, ICU room occupancy, and daily deaths due to COVID-19
\cite{IDPH,Thecityrepo,Thecityny,Nychealth}. We used these data to
perform multi-channel calibration of our model
\cite{Wong2020}, which allows us to infer the underlying time
progression of both $S(t)$ and $R_e(t)$.
The fits for $R_e(S)$ for both cities are shown in
Fig.~\ref{fig:COVID_Re_vs_S}A. In both cases, a sharp drop of $R_e$
that occurred during the early stage of the epidemic is followed by a
more gradual decline. For NYC, there is an extended range over which
$R_e(S)$ has a constant slope in  logarithmic coordinate. This is
consistent with the power law behavior predicted by Eq. \ref{scaling}
with  the slope corresponding to transient immunity factor $\lambda_{\rm eff} =4.5 \pm
0.05$. Chicago exhibits a similar behavior but over a substantially
narrower range of $S$. This reflects the fact that NYC was much harder
hit by the COVID-19 epidemic. Importantly, the range of dates we used
to estimate the immunity factor corresponds to the time interval after
state-mandated Stay-At-Home orders were  imposed,  and before the
mitigation measures   began to be gradually relaxed. The signatures of
the onset of the mitigation and of its  partial relaxation are clearly
visible on both ends of the constant-slope  regime. To examine the
possible effects of variable levels of mitigation on our estimates of
$\lambda_{\rm eff} $ in Fig. \ref{fig:SI_Re_vs_S_mobility} we repeated our
analysis in which $R_e(t)$ was corrected by Google's community mobility
report in these two cities \cite{Goog} (see {\it SI Appendix}). Although the range of
data consistent with the constant slope  shrank somewhat,  our main
conclusion remains unchanged. This provided us with a lower bound
estimate for  the transient  immunity factor: $\lambda_{\rm eff} =4.1\pm 0.1$.

To test the sensitivity of our results to details of the
epidemiological model and  choice of the region we performed an
alternative analysis based on the data reported in
\cite{Juliette2020report}. In that study, the COVID-19 epidemic was
modelled in each of the 50 US states and the District of Columbia.
Because of the differences in population density, level of
urbanization, use of public transport, etc., different states were
characterized by substantially different initial growth rates of the
epidemic, as quantified by the basic reproduction number $R_0$.
Furthermore, the time of arrival of the epidemic also varied a great
deal between individual states,  with states hosting major airline
transportation hubs being among the earliest ones hit by the virus. As
a result of these differences, at any given time the infected fraction
of the population differed significantly across the US
\cite{Juliette2020report}. We use state level estimates of  $R_e(t)$,
$R_0$ and $S(t)$  as reported in   Ref. \cite{Juliette2020report} to
construct the  scatter plot $R_e(t_0)/R_0$ vs $S(t_0)$ shown  in
\ref{fig:COVID_Re_vs_S}, with $t_0$   chosen to be the last reported
date in that study, May 17, 2020.  By performing the linear regression
on these data in logarithmic coordinates, we obtain the fit for the
slope $\lambda_{\rm eff} =5.3\pm 0.6$ and for  $S=1$ intercept   around
$0.54$. In  Fig. \ref{fig:States} (see {\it SI Appendix}), we present
an extended version of this analysis for the 10 hardest-hit states and
the District of Columbia, which takes into account the overall  time
progression of $R_e(t)$ and $S(t)$, and gives similar estimate
$\lambda_{\rm eff} =4.7 \pm 1.5$. Both estimates  of the immunity
factor based on the state data are consistent with our earlier analysis
of NYC and Chicago.  \textcolor{black}{In light of our theoretical
picture, this value of this transient immunity factor,  $\lambda_{\rm
eff} \simeq 4 $, is  set by the  the pace of the first epidemic wave in
the US. As expected, it exceeds our estimate of $\lambda_\infty \approx
2$ associated with persistent heterogeneity  and responsible for the
long-term herd immunity.   }



\begin{figure}[ht]
\centering
\includegraphics[width=0.95\columnwidth]{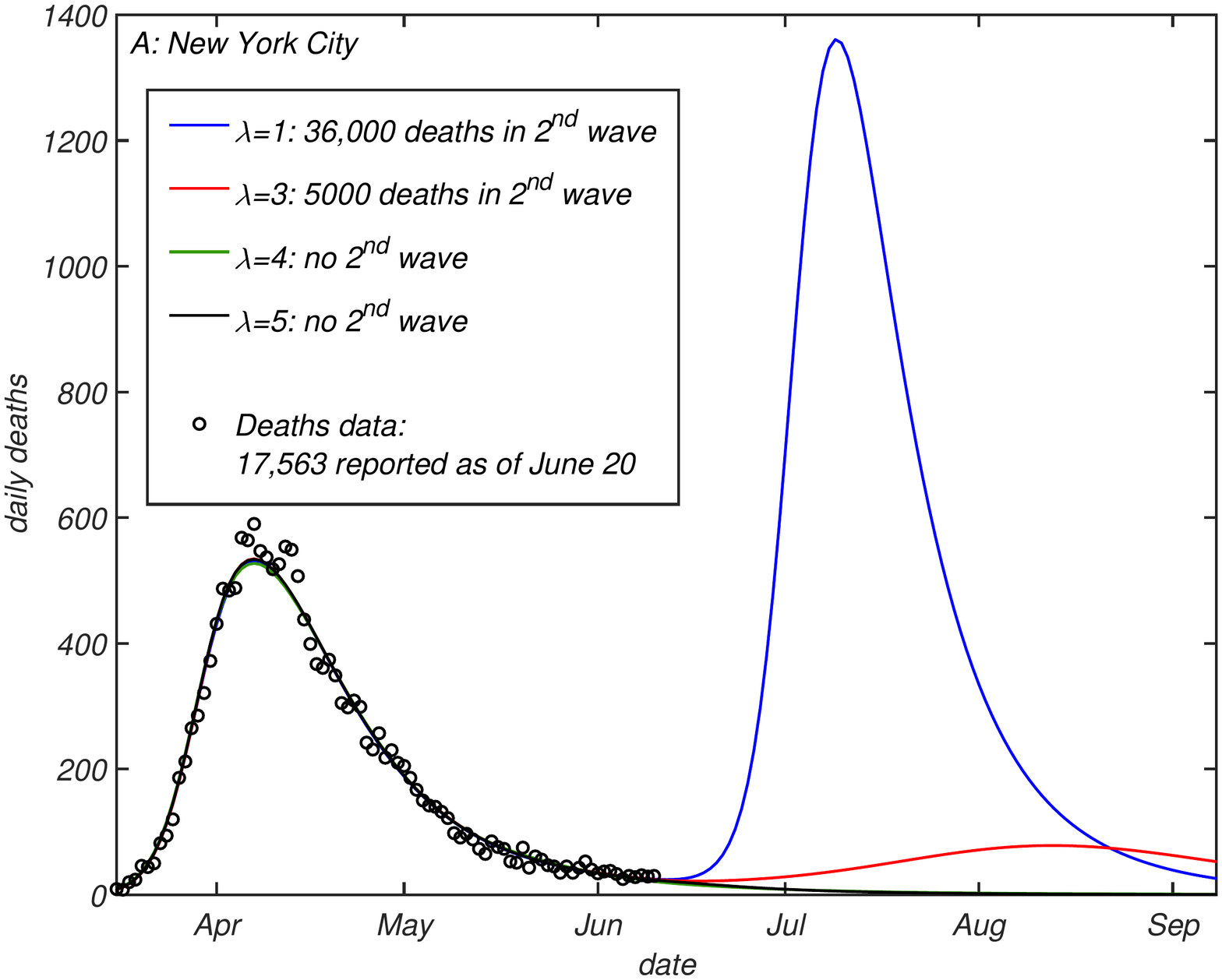}
\includegraphics[width=0.95\columnwidth]{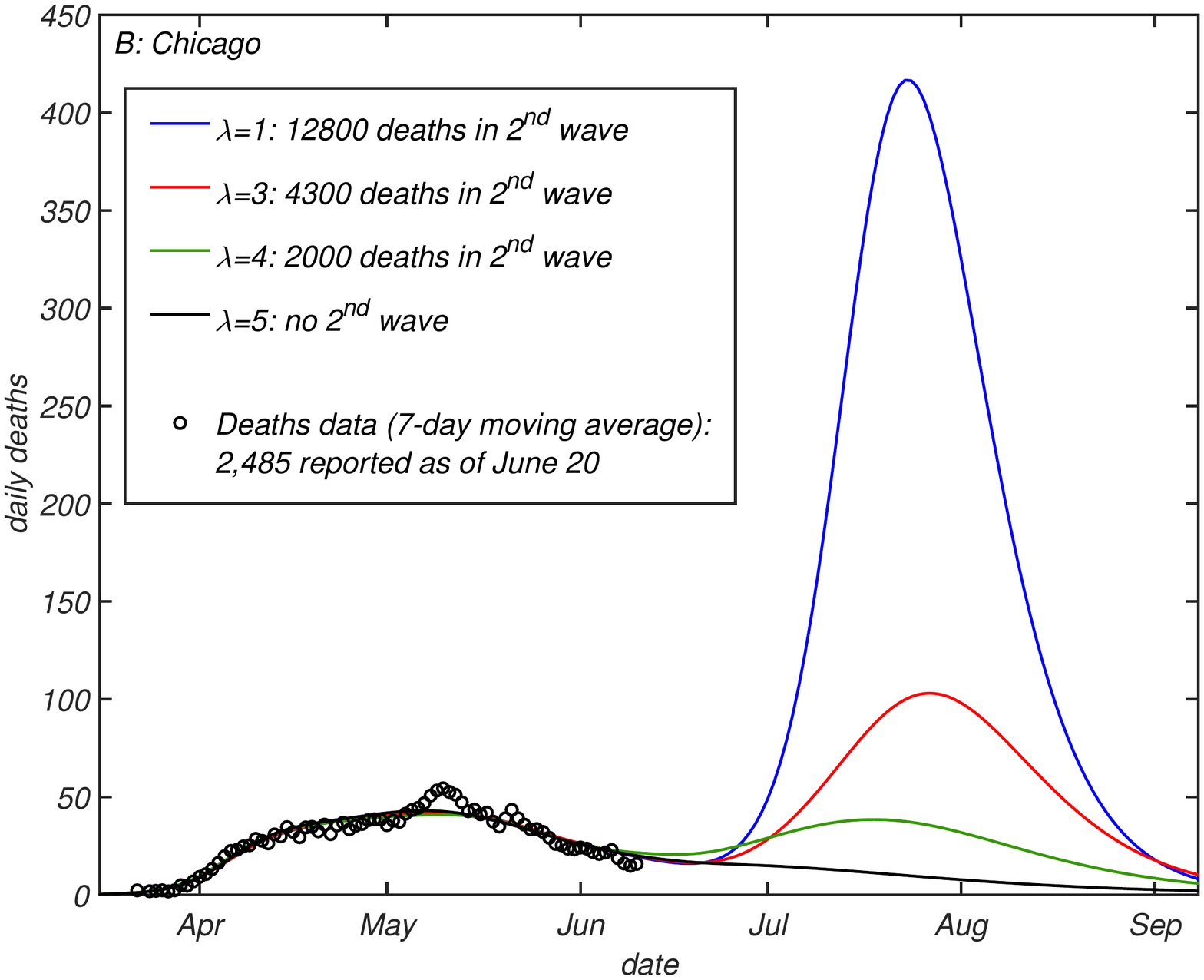}
\caption{Projections of daily deaths under the hypothetical scenario in
which  any mitigation is completely eliminated as of  June 15 2020, for
(A) NYC and (B) Chicago. Different curves correspond to different
values of the transient immunity factor $\lambda_{\rm eff}=1$ (blue), 3
(red), 4 (green) and 5 (black lines). The model described in Ref.
\cite{Wong2020} was fully calibrated on daily deaths (circles), ICU
occupancy and hospitalization data up to the end of May. See {\it SI Appendix}, Figs.
\ref{fig:NYC}-\ref{fig:Chicago} in for additional details,
including confidence intervals. } \label{fig:scenarios}
\end{figure}

We can now incorporate \textcolor{black}{this  transient level of
heterogeneity} into our epidemiological model, and examine how future
projections change as a result of this modification. This is done by
plugging scaling relationships given by Eqs.
(\ref{scaling0})-(\ref{scaling}) into the force of infection  and
incidence rate equations of  the original  model. These equations  are
similar to Eqs. (\ref{attack0})-(\ref{incident}), but also include time
modulation due to the mitigation and a possible seasonal forcing (see
{\it SI Appendix} for more details).  After calibrating the model by
using  the data streams on ICU occupancy, hospitalization and daily
deaths up to the end of May, we explore a hypothetical worst-case
scenario in which any mitigation is completely relaxed as of June 1, in
both Chicago and NYC. In other words, the basic reproduction number
$R_0$ is set back to its value at the initial stage of  the epidemic,
and the only factor limiting the second wave is the partial or full
\textcolor{black}{TCI}, $R_{e}=R_0S^{\lambda}$. The projected daily
deaths for each of the two cities under  this (unrealistically harsh)
scenario are presented in Fig. \ref{fig:scenarios} for various values
of $\lambda$. For both cities, the homogeneous model ($\lambda=1$, blue
lines) predicts a second wave which is larger than the first one with
an additional death toll of around $35,000$ in NYC and $12,800$ in
Chicago. The magnitude of the second wave is greatly reduced by
heterogeneity,  resulting in no second wave in either of the two cities
for $\lambda=5$ (black lines). Even for a modest value $\lambda=3$ (red
lines), which is less than our estimate, the second wave is
dramatically reduced in both NYC and  Chicago (by about $90 \%$ and
$70\%$, respectively).

\section*{\textcolor{black}{Fragility of Transient Collective Immunity}}
\textcolor{black}{ One of the consequences of the bursty nature of social
interactions is that the state of TCI gradually wanes due to changes of
individual social interaction patterns on timescales longer than single
generation interval. This may be viewed as a slow rewiring of social
networks.
}
In the context of the COVID-19 epidemic, individual responses to
mitigation factors such as Stay-at-Home orders may differ across the
population. When mitigation measures are relaxed, individual social
susceptibility $\alpha_s$ inevitably changes. The impact of these
changes on collective immunity depends on whether each person's $\alpha_s$
during and after the mitigation are sufficiently correlated. For
example, the TCI state  would be compromised if people who practiced
strict self-isolation  compensated for it by an above-average
social activity after the first wave of the epidemic has passed.
%
%

\textcolor{black}{ To illustrate the effects of post-mitigation rewiring
of social networks we consider a simple modification of the
heterogeneous model with no persistent heterogeneity ($\alpha=1$ for
everyone) and exponentially distributed instantaneous levels of social
activity $a_i(t)$. This corresponds to $\lambda_{\rm eff}(0)=3$ and
$\lambda_{\rm eff}(\infty)=\lambda_{\infty}=1$. In this model each
individual completely changes the set of his/her social connections at
some time scale $\tau_s$.
These changes
{\it destroy  heterogeneity} giving rise to gradual relaxation of
susceptible fraction $S_{a}$ towards its overall mean value $S$. To
model this, we modify Eq. \ref{generic} to include a simple relaxation term:
\begin{equation}
\dot{S}_{a}=-\alpha S_{a} J - \frac{1}{\tau_s} (S_{a} -S) \quad .
\end{equation}
Epidemiological models with rewiring of underlying social networks have
been studied before \cite{volz2007susceptible} (see
\cite{bansal2010dynamic} for a review), but under a constraint that the
individual level of social activity quantified by network degree is
preserved. In contrast, the dynamics described above stems from the
individual level of social activity $\alpha_s$ changing in time. }

\textcolor{black} {Fig. \ref{fig:rewiring} shows the simulation of SIR
model, where the first wave of the epidemic is mitigated, thereby
reducing the effective reproduction number $R_0=2.5$. During the course
of the mitigation $R_0$ is multiplied by $\mu=0.7$. After the
mitigation measures have been lifted at the end of the first wave, the
population is positioned slightly below the TCI threshold preventing
the immediate start of the second wave. However, gradual rewiring of
the social network with time constant $\tau_s=150$ days ultimately
results in the second and even the third wave of the epidemic (see Fig.
\ref{fig:rewiring}). The inset of this figure shows $R_e(t)$ plotted as
a function of $S(t)$ in this epidemic. Note that each of the waves
follows the power law relationship between $R_e(t) \sim S(t)^{\lambda}$
predicted by Eq. \ref{scaling}. Since constant rewiring eliminates
correlations in  individual social activity  on scales  longer than
$\tau_s$, the epidemic stops after multiple waves bring the total
fraction of infected individuals close to the unmodified (homogeneous)
herd immunity threshold $1/R_0$. Note, however that in this case there
is almost no overshoot and thus the final size of the epidemic is
reduced compared to the case of a purely homogeneous and unmitigated
epidemic.
\begin{figure}[ht]
\centering
\includegraphics[width=0.95\columnwidth]{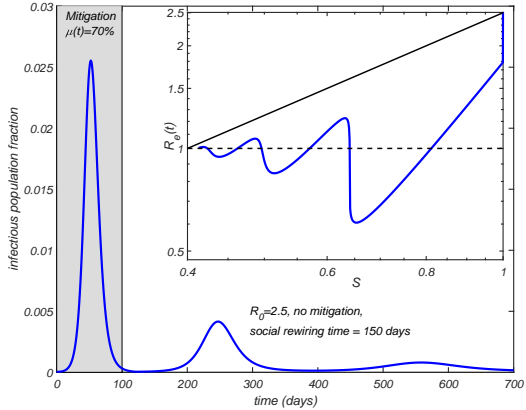}
\caption{\textcolor{black}{Effect of social rewiring on the epidemic
dynamics. The time course of an epidemic in a heterogeneous SIR model
with $R_0=2.5$ and $\lambda=3$. During the first 100 days a mitigation
factor $\mu=0.7$ is applied. Social networks gradually rewire with a
time constant $\tau_s=150$ days. The figure shows multiple waves. The
inset shows $R_e(t)$ plotted as a function of $S(t)$. Solid line shows
the homogeneous limit  reached after multiple waves.}}
\label{fig:rewiring}
\end{figure}
}

\section*{Discussion}

In this work, we have demonstrated \textcolor{black}{how the interplay
between short-term overdispersion and persistent heterogeneity in a
population leads to dramatic changes in epidemic dynamics on multiple
time scales,  transient suppression of the epidemic during its  early
waves, all the way up to the state of long-term  herd immunity. First,
we developed a general approach that allows for the persistent }
heterogeneity  to be easily integrated into a wide class of traditional
epidemiological models in the form of two non-linear functions $R_e(S)$
and $S_e(S)$, both of which are fully determined by the statistics of
individual susceptibilities and infectivities. Furthermore, $R_e(S)$ is
largely defined by a single parameter, the immunity factor $\lambda$,
introduced in our study. Like susceptibility itself, $\lambda$ has two
contributions: biological and social (see Eqs.
(\ref{lambda_b}-\ref{lambda_s})).

\textcolor{black}{ We then expanded our approach to include effects of
time dependence of individual social activity, and in particular of
likely  correlations over the timescale of a single generation
interval.  While our results for purely persistent heterogeneity
confirmed and corroborated   that  HIT  would be suppressed compared to
the homogeneous case, addition of temporal variations led to a dramatic
revision of that simple  narrative. Both persistent heterogeneity and
short-term overdispersion contributions lead first to a slow down of a
fast-paced epidemic, and to its medium-term stabilization. However,
this state of Transient Collective Immunity (TCI) is fragile and does
not constitute long-term herd immunity. HIT is indeed suppressed, but
only due to the persistent heterogeneity. This suppression is
significantly weaker than the initial stabilization responsible for the
TCI state reached after the first wave of a fast-paced epidemic. }

\textcolor{black}{Among other implications of the TCI phenomenon is the
suppression of the so-called overshoot.  Namely, it is well known that
most models predict that an epidemic will not  stop once HIT is passed,
ultimately reaching a significantly larger cumulative attack rate, FSE.
Multiple prior studies
\cite{Newman2002,Moreno2002,Novozhilov,Katriel_FSE_hetro,Ball_hetero_FSE,miller2012note}
have  shown that  FSE would  be  suppressed by persistent
heterogeneity, similarly to HIT. In {\it SI Appendix}, we present a
simple result that unifies several previously studied limiting cases,
and gives an explicit equation for the FSE for the gamma-distributed
susceptibility and variable level of its correlation with infectivity.
However, because of the transient suppression of the early waves of the
epidemic discussed in this work,  the overshoot effect would be much
weaker or essentially  eliminated. For instance, our simple rewiring
model demonstrates how the epidemic, after several waves,  ultimately
reaches HIT level, but does not progress  much beyond it (see Fig.
\ref{fig:rewiring}). The FSE result may still be used, but primarily as
an  estimate for  the size of the first  wave of an (unmitigated)
epidemic. In that case, the transient value of immunity factor
$\lambda_{\rm eff}$ should be assumed. }

By applying our theory to the COVID-19 epidemic we found evidence that
the hardest-hit areas such as New York City, have likely passed
\textcolor{black}{TCI} threshold by the end of the first wave, but are
less likely to have achieved a real long-term herd immunity.  Other
places that had intermediate exposure, such as e.g. Chicago, while
still \textcolor{black}{below} the \textcolor{black}{TCI threshold}, have
their effective reproduction number reduced by a significantly larger
factor than predicted by traditional epidemiological models. This gives
a better chance of suppressing the future waves of the epidemic in
these locations by less disruptive measures than those used during the
first wave, e.g. by using masks, social distancing, contact tracing,
control of potential  super-spreading events,  etc.
\textcolor{black}{However, similar to the case of NYC, transient
stabilization of the COVID-19 epidemic in Chicago will eventually wane.
As for the permanent HIT, although suppressed compared to classical
value, it definitely has not yet been passed in those two locations. }

In a recent study \cite{Herd_Science_2020}, the reduction of HIT due to
heterogeneity has been  illustrated using a toy model. In that model,
$25\%$ of the population was assumed to have their social activity
reduced by $50\%$ compared to a baseline, while  another $25\%$ had
their social activity  elevated twofold. The rest of the population was
assigned the baseline level of activity. According to  Eq.
\ref{lambda_s}, the immunity factor in  that model is  $\lambda=1.54$.
For this immunity factor, Eq. (\ref{eq:HIT}) predicts HIT at
$S_0=64\%$, $55\%$ and $49\%$, for $R_0=2$, $2.5$, and $3$,
respectively. Despite the fact that the model distribution is not
gamma-shaped, these values  are in a very good  agreement  with the
numerical results reported in Ref. \cite{Herd_Science_2020}:
$S_0=62.5\%$, $53.5\%$, and $47.5\%$, respectively.

Thus there is a crucial distinction  between the persistent
heterogeneity, \textcolor{black}{short-term variations correlated over
the time scale of a single generation interval, } and overdispersion in
transmission statistics associated with short-term superspreading
events
\cite{LloydSmith2005,May_superspr_2005,SARS_superspr,Meyers_Newman_SARS_2005,small2006super,Super_Kucharski}.
In our theory, a personal decision to attend a large party or a meeting
would only \textcolor{black}{contribute to persistent heterogeneity}
if it represents a recurring behavioral pattern.
On the other hand, superspreading events are shaped by short-time
variations in individual infectivity (e.g. a person during highly
infectious phase of the disease attending a large gathering). Hence,
the level of heterogeneity inferred from the  analysis of such  events
\cite{LloydSmith2005,SARS_superspr} would be significantly exaggerated
and should not be used to estimate \textcolor{black}{the TCI threshold
and HIT}. Specifically, the statistics of superspreading events is
commonly described by the negative binomial distribution with
dispersion parameter $k$ estimated to be about  $0.1$ for COVID-19
\cite{Super_Kucharski}. According to Ref. \cite{LloydSmith2005}, this
is consistent with the expected value of the individual-level
reproduction number $R_i$ drawn from a gamma distribution with the
shape parameter $k\simeq 0.1$. This distribution has a very high
coefficient of variation,  $CV^2=1/k\simeq 10$. In the case of a
perfect correlation between individual infectivity and susceptibility
$\alpha$, this would result in an unrealistically high estimate of the
immunity factor: $\lambda= 1+2CV^2=1+2/k\simeq 20$. For this reason,
according to our perspective and calculation, the final size of the
COVID-19 epidemic may have been substantially underestimated in Ref.
\cite{Beyond_R0}.  \textcolor{black}{ Similarly, the degree of
heterogeneity assumed in other recent studies
\cite{Herd_Gomes,Aguas_Gomes_Hetero_2020} is considerably larger than
our estimates. Based on our analysis, the value of the immunity factor
$\lambda$ depends on the pace of the epidemic and on the timescale
under consideration. We estimated its long-term value (responsible for
the permanent HIT)  as $\lambda_{\infty}\approx 2$. However, the
transient values are expected to be higher,  especially during the
first several waves of COVID-19 in select locations,  characterized by
large growth rates. Our analysis of the empirical data in NYC and
Chicago indicate that the slowdown of the epidemic dynamics in those
locations was consistent with $\lambda \approx 4$}. In Table
\ref{table:locations} we \textcolor{black}{present our estimates} of the
factor by which $R_e$ is \textcolor{black}{transiently} suppressed as a
result of  depletion of  susceptible population  in selected locations
in the world, as of early June 2020, \textcolor{black}{as well as the
predicted long-term suppression related to acquisition of a partial
herd immunity}.

\begin{table}[ht!]
\caption{ Effects of heterogeneity on  \textcolor{black}{ suppression of the effective reproduction number} $R_e$ in
selected locations. \textcolor{black}{The transient and long-term suppression coefficients $R_e/R_0$ are calculated using
$\lambda = 4$ and $\lambda = 2$, respectively}. Fraction of susceptible
population $S$ as of early June 2020 is estimated from the cumulative
reported death count per capita, assuming the infection fatality rate
(IFR) of  $0.7\%$ \cite{systematic_review_ifr}.  }
\begin{tabular}{lcccc}
\hline
\hline
Location &     Deaths,  & Attack Rate  &      \multicolumn{2}{c}{$R_e$ suppression }\\
 & per 1000 & (1-S) & transient  & long-term \\
\midrule
New York City, USA  \cite{Nychealth}                     &                 $2.1$ & $30\%$ &  $0.24$ & $0.50$ \\
Lombardy, Italy  \cite{Italy}                     &                 $1.7$ & $24\%$ &  $0.33$ & $0.58$\\
London,UK      \cite{London}                &                 $0.9$ & $13\%$ &   $0.58 $ &  $0.75$\\
Chicago, USA \cite{IDPH}                  &                 $0.9$ & $13\%$ &  $0.58 $ &  $0.75$\\
Stockholm, Sweden \cite{Stockholm}                  &                $0.9$ & $13\%$ &  $0.58 $ &  $0.75$\\
\hline\hline
\end{tabular}
\label{table:locations}
\end{table}


Population heterogeneity manifests itself at multiple scales. At the
most coarse-grained level, individual cities or even countries can be
assigned some level of susceptibility and infectivity, which inevitably
vary from one location to another  reflecting differences in population
density and its connectivity to other regions.
Such spatial heterogeneity will result in self-limiting epidemic
dynamics at the global scale. For instance, hard-hit hubs of the global
transportation network such as New York City during the COVID-19
epidemic would gain full or partial \textcolor{black}{TCI} thereby limiting the
spread of infection to other regions during \textcolor{black}{future waves} of
the epidemic. This might be a general mechanism that ultimately limits
the scale of many  pandemics, from the Black Death to the 1918
influenza.

\acknow{
We gratefully acknowledge discussions with
Mark Johnson at Carle Hospital.
The calculations we have performed would have been
impossible without the data kindly provided by the Illinois Department
of Public Health through a Data Use Agreement with Civis Analytics.
This work was supported by the University of Illinois System Office,
the Office of the Vice-Chancellor for Research and Innovation, the
Grainger College of Engineering, and the Department of Physics at the
University of Illinois at Urbana-Champaign. Z.J.W. is supported in part
by the United States Department of Energy Computational Science
Graduate Fellowship, provided under Grant No. DE-FG02-97ER25308. A.E.
acknowledges partial support by NSF CAREER Award No. 1753249. This work
made use of the Illinois Campus Cluster, a computing resource that is
operated by the Illinois Campus Cluster Program (ICCP) in conjunction
with the National Center for Supercomputing Applications (NCSA) and
which is supported by funds from the University of Illinois at
Urbana-Champaign. This research was partially done at, and used
resources of the Center for Functional Nanomaterials, which is a U.S.
DOE Office of Science Facility, at Brookhaven National Laboratory under
Contract No.~DE-SC0012704.}

\showacknow{}


\clearpage
\pagebreak

\renewcommand{\theequation}{S\arabic{equation}}
\setcounter{equation}{0}
\renewcommand{\thefigure}{S\arabic{figure}}
\setcounter{figure}{0}
\setcounter{page}{1}
\onecolumn

\section*{Supporting Information Appendix}
\subsection*{Derivation of quasi-homogeneous model}
\subsubsection*{Age-of-infection model}
We start we the same age-of-infection model as described in the main text, but include additional time-dependent modulation  of the force of
infection :
\begin{equation}
\label{I1}
    J(t)=  \mu(t)\left \langle \int_{0}^{\infty} d\tau R_\alpha K(\tau)j_\alpha(t-\tau)\right \rangle_\alpha
\end{equation}
Here, the  modulation factor $\mu(t)$ can be due (e.g.) to mitigation measures
or seasonal forcing. Due to this modification, Eq. (\ref{Re}) should
be rewritten as follows:
\begin{equation}
\label{Re1}
    S_R(S)\equiv\frac{R_e(t)}{\mu(t)R_0}=\frac{1}{R_0}\int_0^\infty \alpha R_\alpha f(\alpha) e^{-\alpha Z(t)} d\alpha
\end{equation}
Here $R_0=\int_0^\infty \alpha R_\alpha f(\alpha) d\alpha$ is the basic
reproduction number. Now one can write an integral equation for force of
infection which is  formally identical to the one for a homogeneous
case:
\begin{equation}
\label{attack1}
    J(t)=\mu(t)R_0\int_{0}^{\infty} d\tau K(\tau)j^*(t-\tau)
\end{equation}
Here introducing infectivity-weighted incidence rate, $j^*=S_RJ$.
Eq. (\ref{incident}) completes the set of our  quasi-homogeneous equations:
\begin{equation}
\label{incident1}
    \frac{ dS}{d t}=-S_e J
\end{equation}
As discussed in the main text, the inhomogeneity is fully accounted for by non-liner function $S_R(S)$, and variable  effective susceptibility  $\alpha_e(S)$.

\subsubsection*{Compartmentalized SIR/SEIR  models}
The basic  SIR and SIER models can be viewed  as particular   cases of
the age-of infection model discussed above. However, because of their
great importance and wide use,  we present our  construction for a
specific case of  SEIR:
\begin{align}
\dot{S}_\alpha=-\alpha S_\alpha J  \\
\dot{E}_\alpha=\alpha S_\alpha J -\gamma_E E_\alpha \\
\dot{I}_\alpha= \gamma_E E_\alpha -\gamma_I I_\alpha
\end{align}
Here, the force of infection  is $J(t)=\mu(t)\gamma_I\int_0^\infty  R_\alpha I(\alpha) f(\alpha) d \alpha$. We define infectivity-weighted  "Exposed" and "Infectious" fractions as
\begin{align}
    E=\int_0^\infty  R_\alpha E(\alpha) f(\alpha) d \alpha\\
    I=\frac{J}{\gamma_I\mu(t)}=\int_0^\infty  R_\alpha I(\alpha) f(\alpha) d \alpha\\
\end{align}
This leads to a complete description of epidemic dynamics with three
ordinary differential equations, formally equivalent to those for  the  homogeneous case.  The
difference are,  once again,   functions $R_e=\mu(t) S_R(S) R_0$ and
$S_e(S)$:
\begin{align}
\dot{S}=-\mu(t)\gamma_IS_e I  \label{eq:SEIR_S} \\
\dot{E}=R_e(t) \gamma_I  I -\gamma_E E \\
\dot{I}= \gamma_E E -\gamma_I I
\end{align}

\subsection* {Correlation parameter and scaling relationship between infectivity and susceptibility}

Below we consider a model in which biological susceptibility $\alpha_b$
is correlated neither with infectivity nor with social strength
$\alpha_s$ of an individual. On the other hand, both the overall
susceptibility and infectivity are proportional to $\alpha_s$. Let
$f_x$ and $f_y$ be probability density functions (pdfs) of variables
$x\equiv \ln \alpha_s$ and $y\equiv \ln \alpha_b$. It is reasonable to
assume a log-normal distribution for $\alpha_b$, since biological
susceptibility can be modeled  as a product of several random factors
(due to age, gender, genetics, pre-existent conditions, etc).  This
corresponds to a Gaussian form for $f_y$ with variance $\sigma^2$ and
mean $-\sigma^2/2$ (assuming normalization $\langle \alpha_b \rangle=1
$). For a given value of $\alpha$, this translates into Gaussian
distribution of variable $x$ with the same variance, and mean $\ln
\alpha +\sigma^2/2$. This allows us to calculate the average $\alpha_s$
which is proportional to $R_\alpha$:

\begin{equation}
    R_\alpha\sim \langle\alpha_s\rangle \sim \frac{\int f_x(x) \exp \left(x-\frac{(x-\ln \alpha-\sigma^2/2)^2}{2\sigma^2}\right)dx}{\int f_x(x) \exp \left(-\frac{(x-\ln \alpha-\sigma^2/2)^2}{2\sigma^2}\right)dx}
\end{equation}

This integral can be evaluated by the method of steepest descents: for
most pdfs $f_x$ and $f_y$, will be  dominated by the vicinity of  point
$x_0$ defined by the condition $f'(x_0)/f(x_0)=(x_0/\sigma^2-1/2)$. By
expanding $\ln f(x)$ in $x'=x-x_0$, we obtain   $f_x(x')\approx
f(x_\sigma) \exp(r x'-\kappa x'^2/2 ) $, where
$r=f'(x_0)/f(x_0)=x_0/\sigma^2-1/2$ and $\kappa=-f''(x_0)/f(x_0)+r^2$.
After substituting this Gaussian approximation for $f_x$ back into the
above equation, we obtain the scaling relationship between $\alpha$ and
$R_\alpha$
 \begin{equation}
     R_\alpha \sim \exp \left(\frac{(\sigma^2+\ln \alpha)^2-(\ln \alpha)^2}{2\sigma^2(1+\kappa\sigma^2)}\right)\sim \alpha^\chi
    \end{equation}
Here $\chi=1/(1+\kappa\sigma^2)$.

\subsection*{Functions $S_R(S)$ and $S_e(S)$}
According to Eq.(\ref{S_def}), function $S(Z)$ is directly related to
the  moment generating function $M_\alpha$ for pdf $f(\alpha)$
\begin{equation}
    \label{Xz}
    S=\langle e^{-\alpha Z}\rangle_\alpha =M_\alpha(-Z)=1-Z+\frac{\langle\alpha ^2\rangle Z^2}{2}-\frac{\langle\alpha ^3\rangle
    Z^3}{6}+\cdots
\end{equation}
This function also determines the effective fraction of susceptible population $S_e$:
\begin{equation}
    \label{alpha_z}
    S_e=\langle\alpha e^{-\alpha Z}\rangle_\alpha =-\frac{d\ln S}{dZ}
\end{equation}

Remarkably, once the function $S_e(S)$ is found, it completely
determines how $S_R$, and hence $R_e$, behaves in the  limiting cases
of both  the strong and weak correlations:

\begin{equation}
\label{Re_ae}
    S_R= \begin{cases}\langle \alpha e^{-\alpha Z}\rangle_\alpha=-dS/dZ =S_e, & \chi=0 \\
    \frac{1}{\langle \alpha^2\rangle}\frac{dS^2}{dZ^2}=\frac{S_e}{\langle \alpha^2\rangle}\frac{dS_e}{dS}, & \chi=1
    \end{cases}
\end{equation}

\subsection* {Application to specific  distributions of susceptibility}
\subsubsection* {Gamma distribution}
Consider the gamma distribution with $\langle \alpha\rangle=1$ and $CV_\alpha^2=\eta$:
\begin{equation}
\label{gamma}
    f(\alpha)\sim \alpha^{1/\eta-1}\exp(-\alpha/\eta)
\end{equation}
By using Eqs. (\ref{S_def})-(\ref{Re}), we obtain:
\begin{equation}
    S=(1+\eta Z)^{-1/\eta}
\end{equation}
\begin{equation}
    S_e=(1+\eta Z)^{-1/\eta-1}=S^{1+\eta}
\end{equation}
\begin{equation}
    S_R=(1+\eta Z)^{-(1+(\chi+1)/\eta)}=S^\lambda
\end{equation}
This leads to the scaling relationship $R_e=R_0S^\lambda$, Eq. (\ref{scaling}).

\subsubsection*{Truncated power law distribution}
We now consider power law distributed $\alpha$, $f(\alpha)\sim
1/\alpha^{1+s} $ ($s>0$), with upper and lower  cut-offs, $\epsilon
\alpha_+$ and  $\alpha_+$, respectively. If the upper cut-off is
implemented as an exponential factor $\exp(-\alpha/\alpha_+)$, we recover
the functional form identical to the gamma distribution, Eq.
(\ref{gamma}) discussed above, but with  negative values of the shape
factor:
\begin{equation}
    f(\alpha)=\frac{\alpha_+^{q-1}\exp(-\alpha/\alpha_+)}{\alpha^{q}\Gamma(1-q,\epsilon)}
\end{equation}
Due to the normalization $\langle\alpha\rangle=1$,
\begin{equation}
   \alpha_+=\frac{\Gamma(1-q,\epsilon)}{\Gamma(2-q,\epsilon)}.
\end{equation}
In the case of gamma distribution, the coefficient of variation
$CV_\alpha$ would completely determine the overall shape of pdf. For
power law with exponent $1\le q\le 3$, the  value of $\eta=CV^2$ sets
the dynamic range  the between upper and lower cut-offs, i.e. the parameter
$\epsilon$:
\begin{equation}
    1+\eta=\langle\alpha^2\rangle=\frac{\Gamma(1-q,\epsilon)\Gamma(3-q,\epsilon)}{\Gamma(2-q,\epsilon)^2}
    \end{equation}

By using Eq. (\ref{S_def})-(\ref{Re}), we  can obtain exact results for $S$ $S_R$ in terms of  $Z$:
\begin{equation}
    S =\frac{\Gamma(1-q,\epsilon(1+\alpha_+Z))}{\Gamma(1-q,\epsilon)(1+\alpha_+Z)^{1-q}}
\end{equation}
\begin{equation}
S_R=\frac{\Gamma(\nu,\epsilon(1+\alpha_+Z))}{\Gamma(\nu,\epsilon)(1+\alpha_+Z)^\nu}
\end{equation}
Here $\nu=2+\chi-q$. The resulting function $R_e/R_0=S_R(S)$ is shown in Fig. \ref{fig:banana} for several values of the exponent $q$.

For  $\chi=0$, the overall function $S_R(S)=S_e(S)$ can be very well
fitted by an empirical analytic formula that depends only on
$\lambda_0=1+CV_\alpha^2$ and an additional shape parameter
$\Delta_\lambda=CV_\alpha (\gamma_\alpha-2CV_\alpha)$:
 \begin{equation}
   S_e(S)\approx \frac{S}{\left(1+\Delta_\lambda(1-S)\right)^{(\lambda_0-1)/\Delta_\lambda}}
    \end{equation}
 According to Eq.  (\ref{Re_ae}),  this function completely defines behavior of $S_R$ in both limits of  the  weak and strong correlation regimes :
\begin{equation}
    S_R\approx \frac{\left(1+(\Delta_\chi-1)(1-S)\right) S}{\left(1+\Delta_\lambda(1-S)\right)^{(\lambda-\Delta_\chi)/\Delta_\lambda}}
\end{equation}
 Here $\Delta_\chi=(\Delta_\lambda+1)/\lambda_0$, and $\lambda=\lambda_1$ for $\chi=1$. For $\chi=0$,  $\delta_\chi$ has to be set to $1$.


\subsubsection* {Log-normal distribution} The log-normal distribution is a
very natural candidate to describe statistics of $\alpha$. It
universally emerges for multiplicative random processes. Transmission
of an infection involves a complex chain of random events, both social
and biological, which can be conceptualized as such multiplicative
process. For instance, it may depend on  how likely a given person
would be involved in a potential superspreading event, how likely that
person would have a close contact with a potential infector, what would
be the duration of their contact, how effective the individual immune
system is in preventing and suppressing the infection.

\begin{figure}[ht!]
\centerline{\includegraphics[width=1\paperwidth]{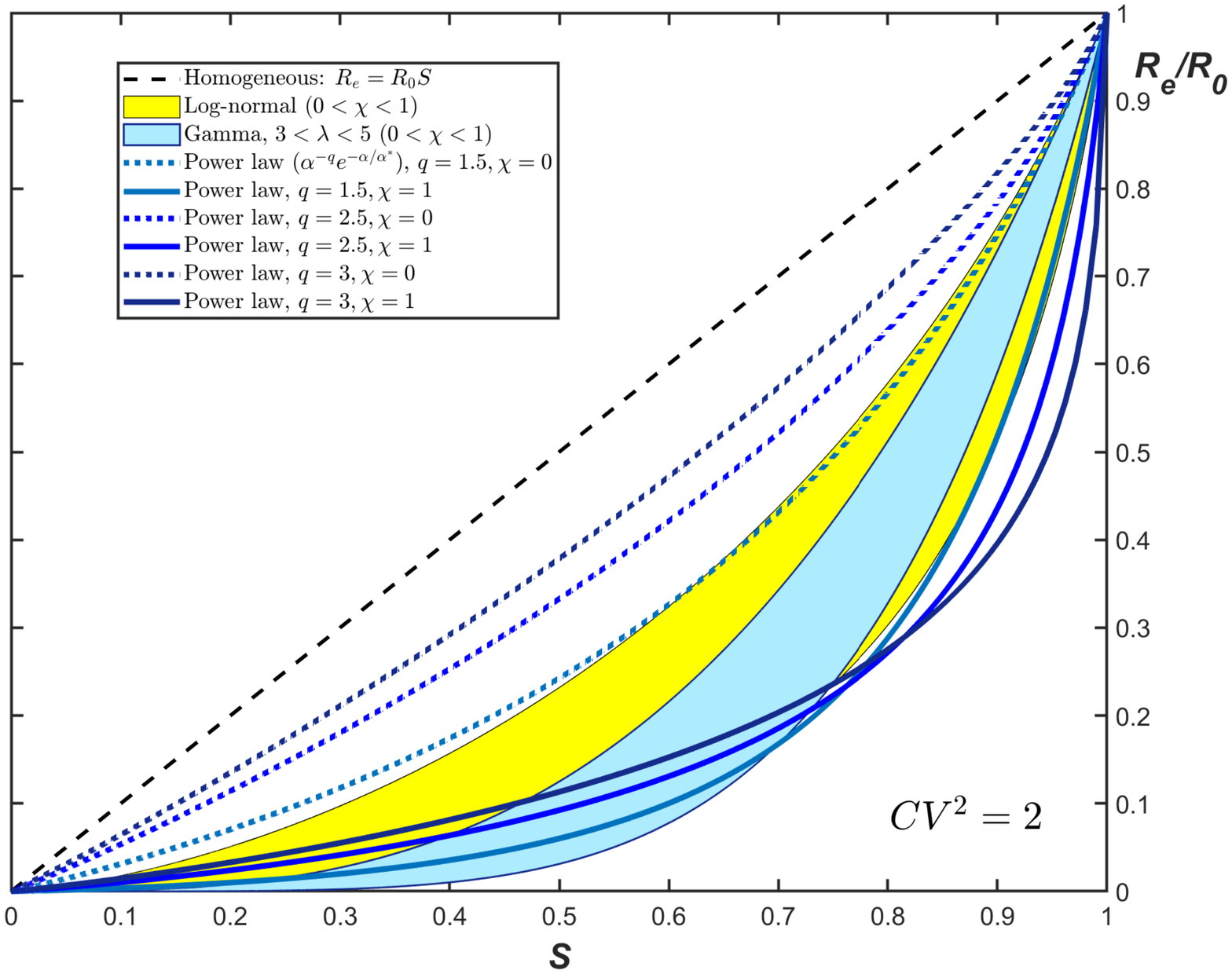}}
\caption{$R_e/R_0$ vs $S$ dependence for three different families of probability distribution $f(\alpha)$: Gamma (light blue), truncated power law (dashed lines), and log-normal (yellow). Different curves correspond to the same value of the coefficient of variation $CV^2_\alpha=2$, and two limiting values (0 and 1) of the correlation parameter $\chi$.}
\label{fig:banana}
\end{figure}

For the log-normal distribution, the initial drop in $R_e$ according to
Eq. (\ref{lambda2}), is noticeably faster than for a gamma distribution:
$\lambda=(1+CV_\alpha^2)(1+\chi CV_\alpha^2)$. However, the initial
linear regime is also  much narrower. Figure \ref{fig:banana} shows the
dependence $R_e(S)$ for the log-normal distribution alongside with the
above results for gamma and power law distributions computed  for the
same values of CV (specifically, $CV^2_\alpha=2$). As one can see from
these plots, despite a stronger effect of heterogeneity at the early
stage, the curves generated by log-normal distribution approach $R_e=0$
significantly  slower  than those corresponding to the gamma
distribution. Note that the overall behavior of $R_e(S)$  generated by
the log-normal distribution closely matches the one obtained for the
power law distribution with a certain scaling exponent  $q$. This
exponent would depend on $CV$ and should approach $1$ in the limit of
sufficiently wide distribution when the log-normal pdf asymptotically
approaches a power law $1/\alpha$ with upper and lower cut-offs.

\subsection*{Final Size of Epidemic}

Here we  derive   a simple  result for FSE  in a population with a
persistent heterogeneity. To do this, we integrate Eq. (\ref{attack0})
over time $t$. This yields a relation $Z_\infty=\int_0^{\infty}
R_e(t)J(t)dt =\int_{S_\infty}^{1}R_e(S)dS/S_e(S)$ for the final value
of $Z$ when the epidemic has run its course, and this in turn can
conveniently be expressed in terms of the fraction of the susceptible
population, $S_\infty$:
\begin{equation}
\label{FSE1}
    S_\infty = M_\alpha\left(-\int_{S_\infty}^1 \frac{R_e(S) dS}{S_e(S)} \right)
\end{equation}
This equation is valid  for an arbitrary distribution of $\alpha$,
arbitrary correlation between susceptibility and infectivity, and for
any statistics  of the  generation interval. \textcolor{black}{This
result  can be also obtained as a solution to a general integral
equation derived in Ref. \cite{miller2012note} for the well-mixed
case}.  Eq. \ref{FSE1} combines and generalizes several well-known
results:  (i) in  the weak  correlation limit ($R_\alpha=R_0$),  when
the integral in the r.h.s. is equal to $R_0(1-S_\infty)$,
Eq.(\ref{FSE1}) reproduces results of Refs.
\cite{Novozhilov,Katriel_FSE_hetro,Ball_hetero_FSE,miller2012note},
(ii) in the opposite limit of a strong correlation ($R_\alpha\sim
\alpha$),   the integration gives $R_0(1-S_e(S_\infty))/\langle
\alpha^2\rangle$,  and one recovers the result for the FSE on a network
\cite{Newman2002,Moreno2002,miller2012note}.

For the case of
gamma-distributed persistent  susceptibility  Eq.
(\ref{FSE1}),  gives:
\begin{equation}
\label{FSE2}
S_{\infty}=\left(1+\frac{R_0 \eta\left(1-S_{\infty}^{\lambda-\eta}\right)}{\lambda-\eta}\right)^{-1/\eta}
\end{equation}
\textcolor{black}{It should be emphasized however that  this result is
of limited relevance to more realistic situations. Even if one assumes
no government-imposed mitigation or societal response to the epidemic,
the case of fully persistent heterogeneity is just  an approximation.
As we demonstrate in our paper,  short-term correlations of
time-dependent individual susceptibilities and infectivities lead to
transient stabilization of a fast-pacing epidemic. Because of this
effect, Eq.(\ref{FSE1})-Eq.(\ref{FSE2}) should be interpreted as an
estimate of the size of the first wave rather than the actual FSE.  }

\subsection* {\textcolor{black}{Path-integral theory of epidemic with time-dependent heterogeneity}}

\textcolor{black}{Here we present a generalization of the theory
developed in the previous section that incorporates the effects of time
variations of individual susceptibilities and infectivities, as well as
temporal correlations between them. Since these fast variations are
primarily caused by bursty dynamics of social interactions, and since
heterogeneous biological susceptibility appears subdominant in the
context of COVID-19, we set $\alpha_b=1$ for all individuals, so that
$\alpha$ has purely social origin. Let   $a_i(t)=\alpha_i+\delta
a_i(t)$ be the time-dependent susceptibility of a  person.  Because of
the social nature of $ a(t)$, one's individual infectivity  is also
proportional to it at any given time: $\beta_i(t)=R\cdot K(\tau)
a_i(t)$.   As before,  $\tau$ is time from infection, $K(\tau)$ is  the
pdf of generation intervals. Accordingly $R$ is individual
reproductive number of an "average" person with social activity
$a_i(t)=1$, in the fully susceptible population.   The state of an
individual is described by a step function $s_i(t)$ which is $1$ as
long as the person is susceptible,  and  turns to $0$ at the moment of
infection. The time evolution of the epidemic follows a stochastic
generalization of  Eqs. (\ref{generic})-(\ref{I}):
\begin{align}
    &{\rm E}\left [{\dot{s}_i(t)}\right]=-a_i(t) s_i(t-0) J(t) \label{eq:prob}\\
& J(t)=-\int_0^\infty R \cdot K(\tau)\overline{ a_i(t)\dot{s}_i(t-\tau)} d \tau\label{eq:J_new}
\end{align}
Here bar $\bar{\dots}$ represents averaging over  individual members of
population (indexed by $i$), in contrast with $\langle \dots \rangle$,
averaging over all subgroups with various values of persistent
heterogeneity $\alpha$.   ${\rm E}[\dots]$ stands for expected value.    }

\textcolor{black}{The overall quasi-homogeneous description given by
Eqs.(\ref{incident})-(\ref{attack0}), remains valid. It is  obtained by averaging Eqs. (\ref{eq:prob})-(\ref{eq:J_new}) over the
entire population. However, in contrast to the case of persistent
heterogeneity,   variables  $S(t)$, $S_e(t)$ and $R_e$ are no longer
connected to each other by a simple functional relationship. To  relate
them we first note that the average probability that an individual is
still susceptible at time $t$ is given by
$E[s_i(t)]=\exp\left(-\int_{-\infty}^{t}J(t') a_i(t')dt'\right)$.
Therefore,
\begin{equation}
\label{path_int}
S(t,[J(t')])\equiv\overline{ s_i(t)}=\overline{\exp\left[-\int_{-\infty}^{t}J(t') a_i(t')dt'\right]}
\end{equation}
In other words, $S$ becomes a functional over the set of all possible
epidemic trajectories $J(t)$. It still has the structure of a moment
generating function for the field $a_i(t)$, and thus is a direct analogue
of the partition function broadly used in statistical physics,
stochastic calculus, and  field theory.  The specific form of this
functional depends on probabilities assigned to different individual
trajectories  $a_i(t)$. As a natural generalization of the case of
persistent heterogeneity, $S_e(t)$ and $R_e(t)$ can be obtained as,
respectively, the first and the second variations of the functional $S$
over $J(t)$:
\begin{align}
S_e(t)&=\overline{ a_i(t) s_i(t)}=-\frac{\delta S(t,[J(t')])}{\delta J(t)} \label{eq:Se_path}\\
R_e(t)&=R\overline{a_i^2(t) s_i(t)}=R\frac{\delta^2 S(t,[J(t')])}{\delta J(t)^2}\label{eq:Re_path}
\end{align}
For the sake of simplicity, in deriving Eq.(\ref{eq:Re_path}) we
assumed $\alpha_i(t)$ to be smoothed over the  timescale of a single
generation interval. As a result,  $\int_0^\infty \overline{a_i(t)
a_i(t-\tau)} K(\tau)d\tau\approx \overline{ a_i^2(t)}$. By applying
Eq.(\ref{eq:Re_path}) to the initial state of fully susceptible
population we obtain the result for $R_0$, Eq. (\ref{R0}):
    \begin{equation}
    R_0=R\left(\langle \alpha^2\rangle+\overline{\delta a_i^2}\right)
\end{equation}}

\textcolor{black} {At the early stages of epidemic $E[s_i(t)]\approx
1-\int_{-\infty}^t a_i(t')J(t')dt'$ for the entire population. After
substituting this expression for $s_i(t)$ to Eq.(\ref{eq:Re_path}) one
obtains  a generalization of our previous result for the initial
suppression of $R_e$, Eqs.(\ref{lambda1})-(\ref{lambda2}):
\begin{align}
\label{Re-lambda_SI_new}
R_e(t)&\approx R_0\left(1 -\int_{0}^{\infty}\Lambda(t,t')J(t-t')dt'\right) \\
\label{lambda_kernel}
\Lambda(t,t')&=\frac{\overline{\tilde{\alpha}^2_i(t)a_i(t-t')}}{\langle \alpha^2\rangle +\overline{ \delta a_i^2}}=\lambda_\infty +\delta\lambda(t,t')
\end{align}
Here $\lambda_\infty=\Lambda(\infty)$ and $\delta\lambda(t,t')$ are the
constant and time-dependent contributions to "immunity kernel"
$\Lambda(t,t')$ which are discussed in the main text.}

\textcolor{black}{To obtain a corrected result for HIT, we assume a very
slow progression of the epidemic (e.g. due to a gradual relaxation of
the level of mitigation). In this case, any intermediate-term
correlations between time dependent variations $\delta \alpha_i(t)$
become negligible, and we largely recover the formalism developed for
pure persistent heterogeneity. Under the same assumption that was used
for the estimate of $\lambda_\infty$ in the main text (i.e.
$\overline{\delta a_i^2}\sim \alpha_i$), the only modification that
needs to be done  is to replace the term $\alpha R_\alpha$ in Eq.
(\ref{Re}) with $(\chi^* \alpha^2  + (1-\chi^*)\alpha)   R_0$ (where
$\chi^*=\langle \alpha^2\rangle/\overline{a_i^2}$). This gives an
expression for $R_e(S)$ in terms of two earlier  results,
$R_e^{(\chi)}(S)$ (for $\chi=0$ and $1$, respectively):
\begin{equation}
\frac{R_e}{R_0}=\chi^* R_e^{(1)}+(1-\chi^*) R_e^{(0)}\approx  S^{\lambda_\infty}
\end{equation}
Here $\lambda_\infty=1+(1+\chi^*)\eta$. To obtain this result,  we
substituted the scaling functions for gamma-distributed persistent
heterogeneity, $R_e(S)=R_0S^{\lambda}$, with $\lambda=1+(1+\chi)\eta$,
as given by  Eq.(\ref{scaling}).  }
\clearpage
\pagebreak
\pagebreak 
\begin{figure}[ht!]
\centerline{\includegraphics[width=1\paperwidth]{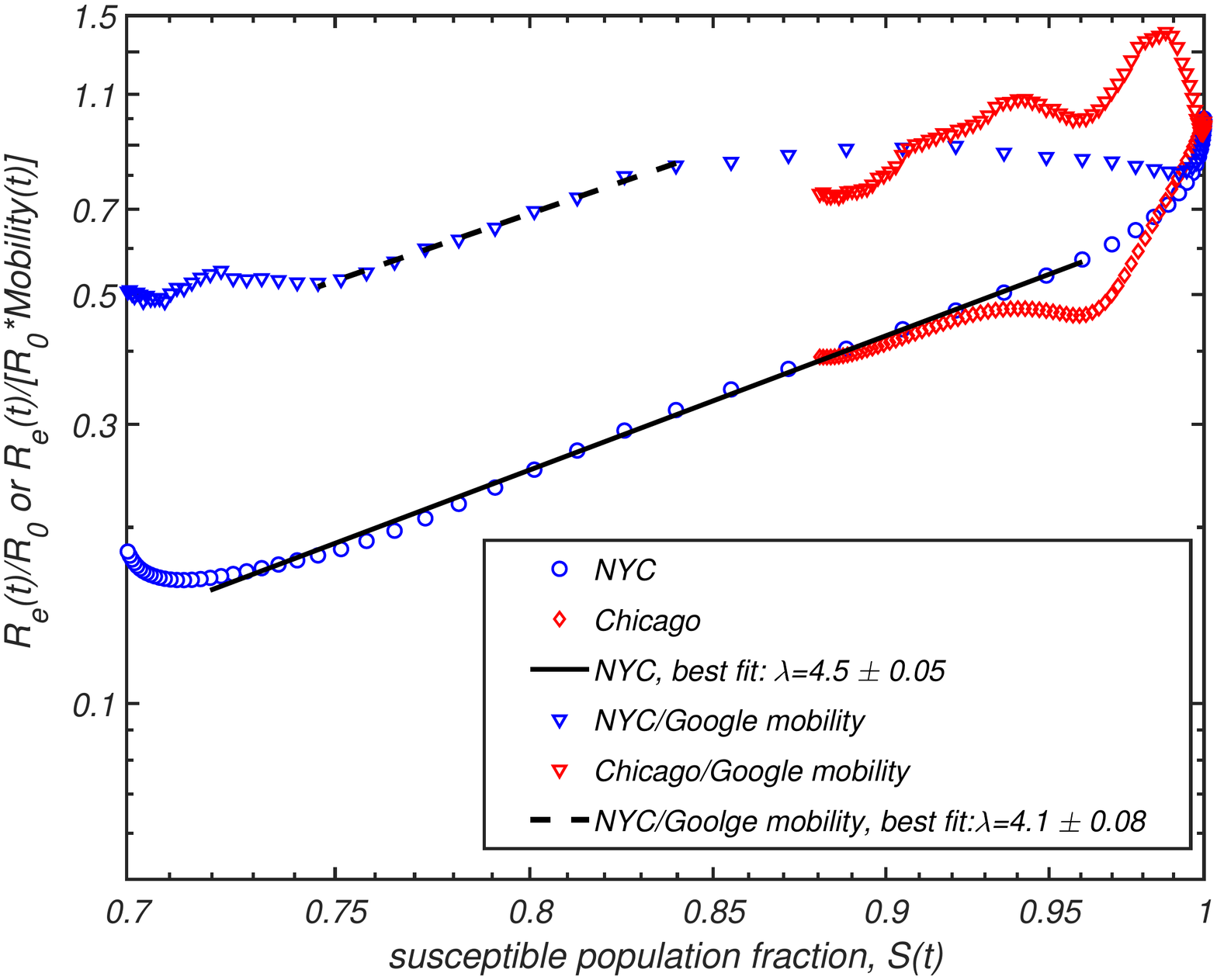}}
\caption{Exploration of effect of mobility on data presented in Figure \ref{fig:COVID_Re_vs_S}(A). Triangles represent data points for NYC and Chicago  with $R_e(t)/R_0$ corrected by a mobility factor calculated from  Google  community mobility report, Ref. \cite{Goog}.
We compute the mobility for NYC using average mobility of its five counties: New York county, Bronx county, Kings county, Richmond county, and Queens county, weighted by their population fraction.
}
\label{fig:SI_Re_vs_S_mobility}
\end{figure}
\pagebreak 
\begin{figure}[ht!]
\centerline{\includegraphics[width=1\paperwidth]{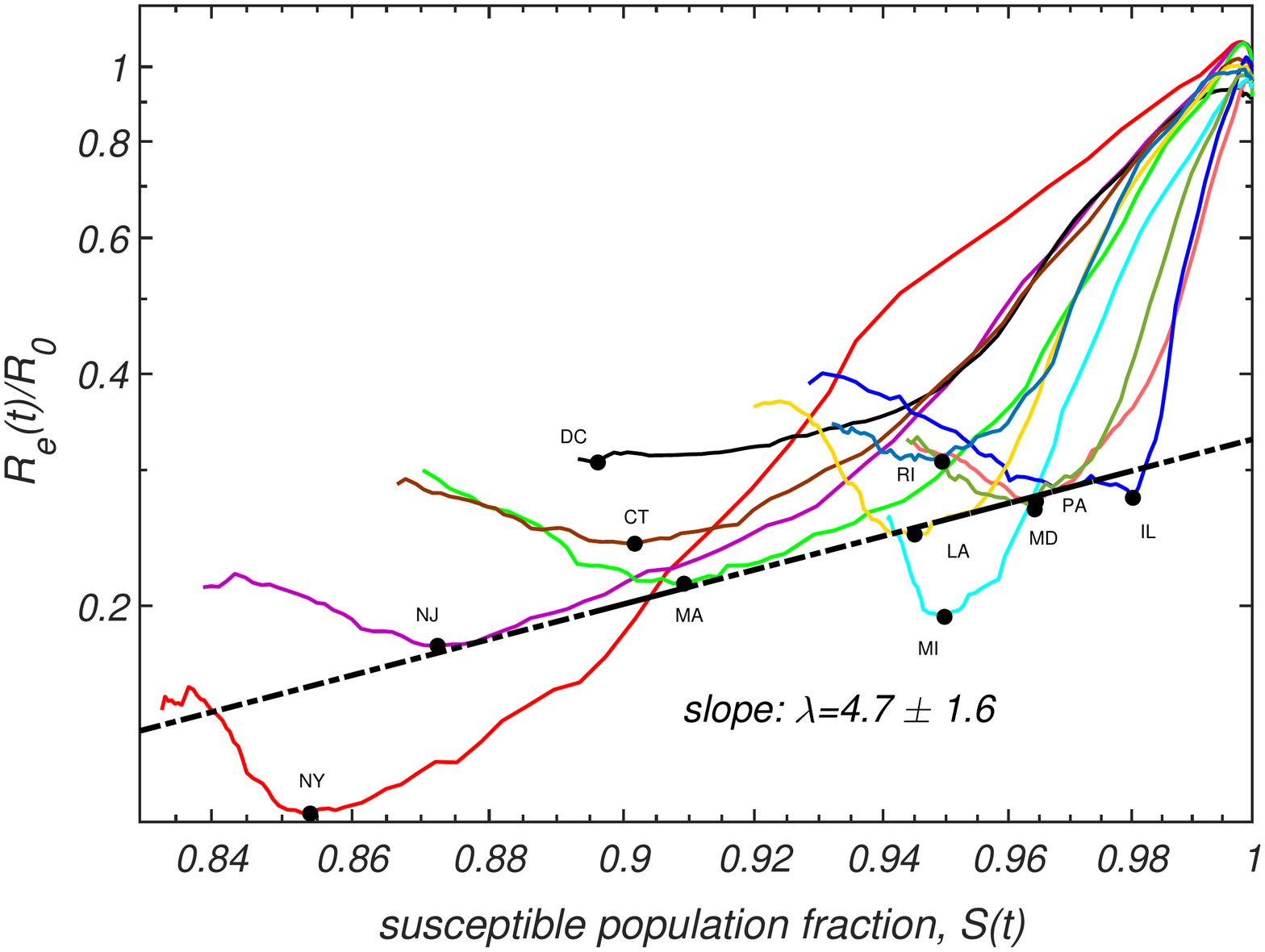}}
\caption{Time progressions of $Re(t)/R_0$ and $S(t)$  for the hardest-hit US states and DC, as reported in Ref. \cite{Juliette2020report}. Black dots correspond to absolute minima of transmission and population susceptible fractions. The dashed line with slope $\lambda=4.7\pm 1.6$ is the best power law fit through these black dots.}
\label{fig:States}
\end{figure}

\pagebreak
\begin{figure*}[ht!]
\centerline{\includegraphics[width=1\paperwidth]{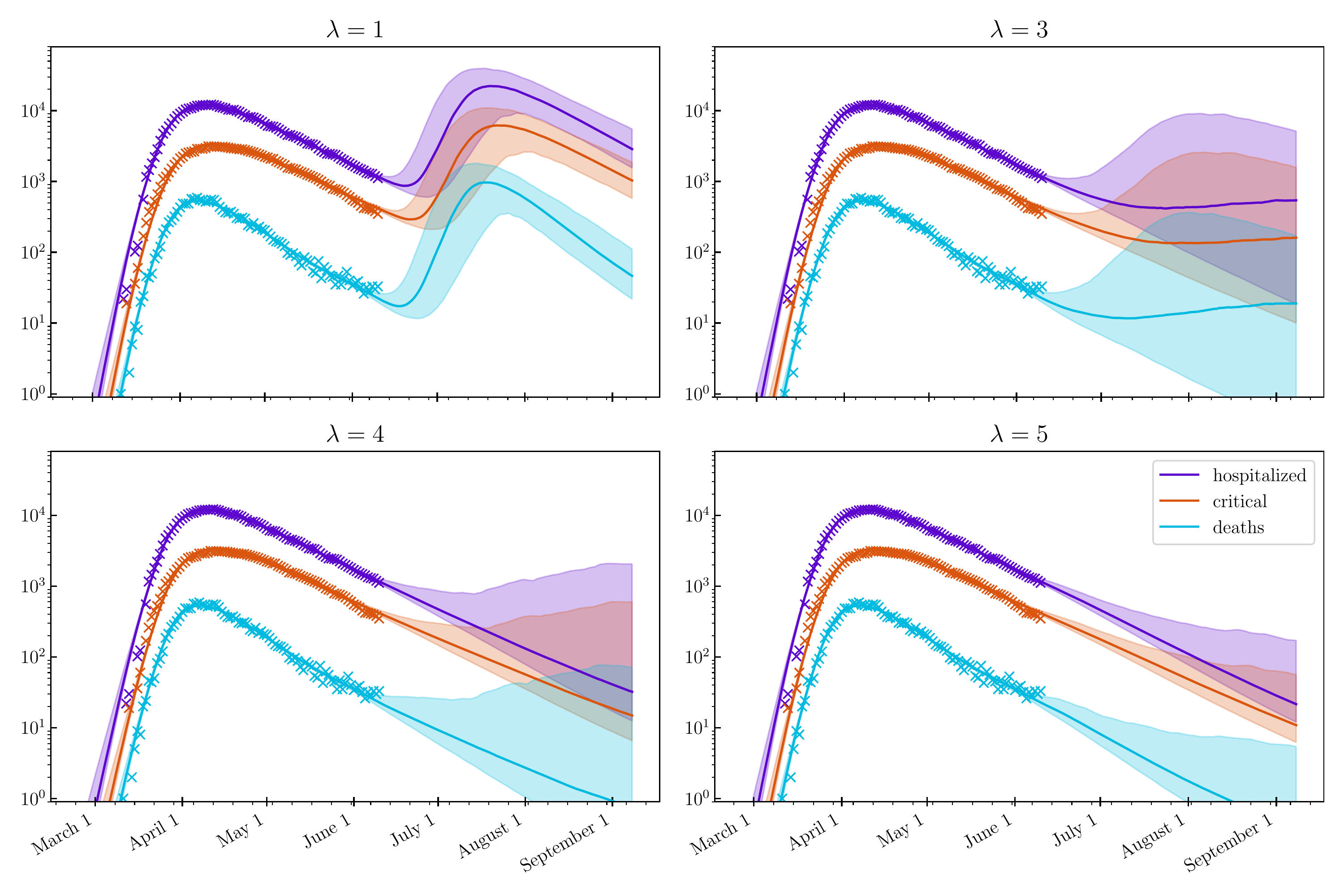}}
\caption{Hospitalization, ICU occupancy and daily deaths in NYC modeled under  hypothetical scenario when any mitigation is completely eliminated  as of  Jun 15 2020, for various values of $\lambda$.  Model described in Ref. \cite{Wong2020} is calibrated on data from  Ref.\cite{IDPH}, up to June 10,  2020 (shown as crosses).  $95\%$ confidence intervals are indicated.    }
\label{fig:NYC}
\end{figure*}
\pagebreak 
\begin{figure*}[ht!]
\centerline{\includegraphics[width=1\paperwidth]{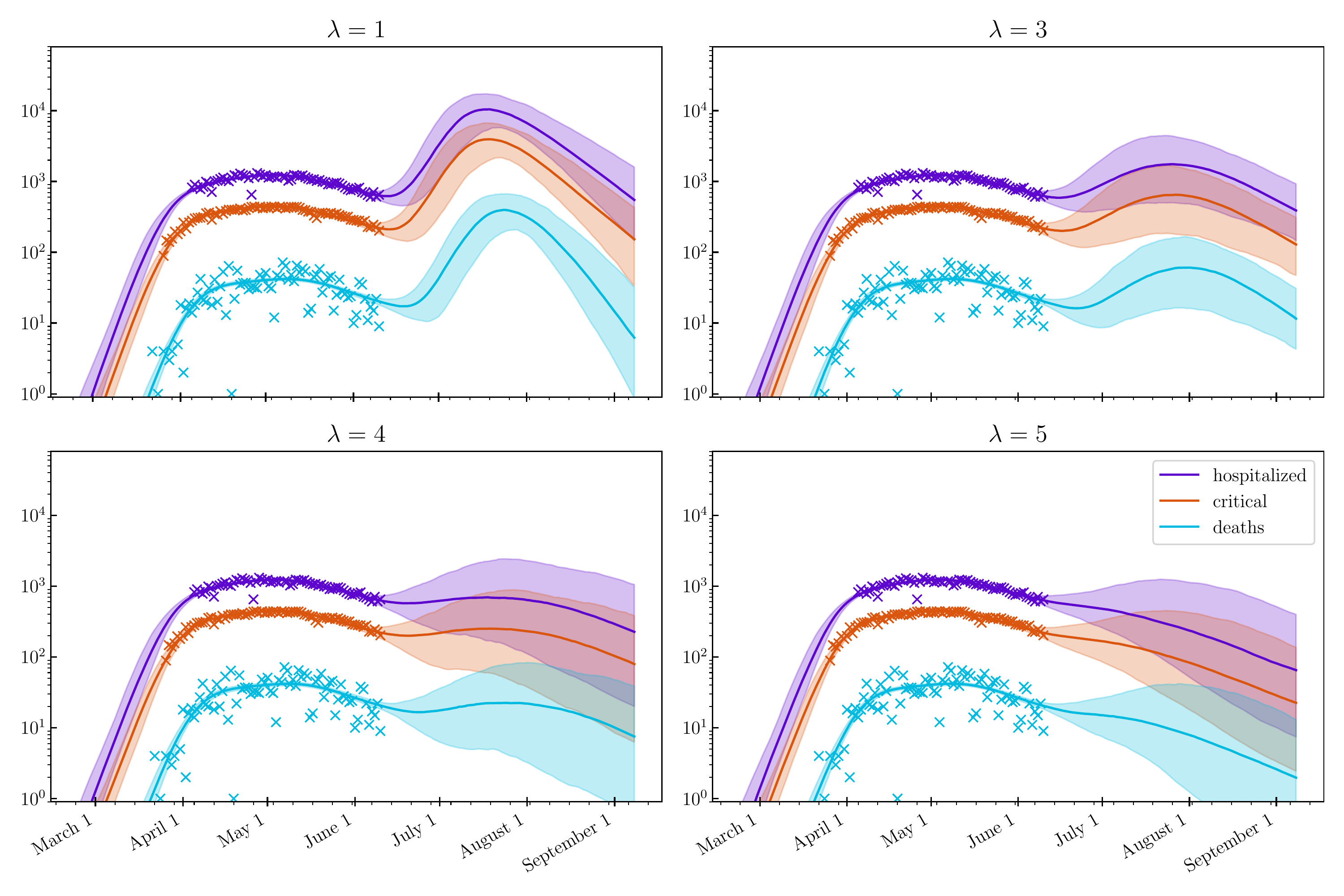}}
\caption{Hospitalization, ICU occupancy and daily deaths in Chicago modeled under  hypothetical scenario when any mitigation is completely eliminated  as of  Jun 15 2020, for various values of $\lambda$.  Model described in Ref. \cite{Wong2020} is calibrated on data from  Ref.\cite{IDPH}, up to June 10,  2020 (shown as crosses).  $95\%$ confidence intervals are indicated. }
\label{fig:Chicago}
\end{figure*}
\pagebreak
\end{document}